\documentclass[preprint,secnumarabic,showpacs,amsmath,amssymb,floatfix,superscriptaddress,longbibliography,aps,prb]{revtex4-1}

\usepackage{graphicx}
\usepackage{color}%
\usepackage{bm}
\usepackage{paralist}
\usepackage[utf8]{inputenc}
\newlength{\alphabet}
\usepackage{hyperref} 
\usepackage{verbatim} 
\usepackage{amsmath} 
\usepackage{amsfonts}
\usepackage{amssymb}
\usepackage{tabularx}  
\usepackage{units} 
\usepackage{physics} 
\usepackage{bbold} 
\usepackage{bbm} 
\usepackage{slashed}
\usepackage{float}
\usepackage{txfonts}
\usepackage[english]{babel}
\usepackage{caption}
\usepackage{soul}
\usepackage[authormarkup=none]{changes}
\definechangesauthor[name=C. Pfleiderer, color=blue]{CP}
\definechangesauthor[name=M.A. Wilde, color=red]{MW}
\definechangesauthor[name=V. Leeb, color=magenta]{VL}
\definechangesauthor[name=J. Knolle, color=green]{JK}
\definechangesauthor[name=N. Huber, color=orange]{NH}

\renewcommand{\v}[1]{\bm{#1}} 

\usepackage{tikz} 
\usetikzlibrary{arrows.meta}
\newcommand{\tstar}[5]{
	\pgfmathsetmacro{\starangle}{360/#3}
	\draw[#5] (#4:#1)
	\foreach \x in {1,...,#3}
	{--(#4+\x*\starangle-\starangle/2:#2) -- (#4+\x*\starangle:#1)}
	-- cycle;}

\newcommand{\rxx}{$\rho_{xx}$}
\newcommand{\ryx}{$\rho_{xy}$}

\newcommand{\tilrxx}{$\tilde{\rho}_{xx}$}
\newcommand{\tilryx}{$\tilde{\rho}_{xy}$}

\begin{document}

\renewcommand{\floatpagefraction}{0.5}
\bibliographystyle{nature}


\title{Quantum Oscillations of the Quasiparticle Lifetime in a Metal}


\author{Nico Huber\footnote{These authors contributed equally.}}
\affiliation{Technical University of Munich; TUM School of Natural Sciences, Department of Physics, 85748 Garching, Germany}

\author{Valentin Leeb$^{\rm a}$}
\affiliation{Technical University of Munich; TUM School of Natural Sciences, Department of Physics, 85748 Garching, Germany}
\affiliation{Munich Center for Quantum Science and Technology (MCQST), 80799 Munich, Germany}

\author{Andreas Bauer}%
\affiliation{Technical University of Munich; TUM School of Natural Sciences, Department of Physics, 85748 Garching, Germany}
\affiliation{Technical University of Munich; Centre for Quantum Engineering (ZQE), 85748 Garching, Germany}

\author{Georg Benka}%
\affiliation{Technical University of Munich; TUM School of Natural Sciences, Department of Physics, 85748 Garching, Germany}

\author{Johannes Knolle\footnote{johannes.knolle@tum.de}}
\affiliation{Technical University of Munich; TUM School of Natural Sciences, Department of Physics, 85748 Garching, Germany}
\affiliation{Technical University of Munich; Munich Center for Quantum Science and Technology (MCQST), 85748 Garching, Germany}
\affiliation{Blackett Laboratory, Imperial College London, London SW7 2AZ, United Kingdom}

\author{Christian Pfleiderer\footnote{christian.pfleiderer@tum.de}}
\affiliation{Technical University of Munich; TUM School of Natural Sciences, Department of Physics, 85748 Garching, Germany}
\affiliation{Technical University of Munich; Munich Center for Quantum Science and Technology (MCQST), 85748 Garching, Germany}
\affiliation{Technical University of Munich; Centre for Quantum Engineering (ZQE), 85748 Garching, Germany}

\author{Marc A. Wilde\footnote{marc.wilde@tum.de}}
\affiliation{Technical University of Munich; TUM School of Natural Sciences, Department of Physics, 85748 Garching, Germany}
\affiliation{Technical University of Munich; Centre for Quantum Engineering (ZQE), 85748 Garching, Germany}

\date{\today}

\maketitle


\noindent \textbf{
Following nearly a century of research, it remains a puzzle that the low-lying excitations of metals are remarkably well explained by effective single-particle theories of non-interacting bands~\cite{1961_Luttinger_PR, 1970_Engelsberg_PRB, 1989_Wasserman_JPCM, 1987_Taillefer_JMMM}. The abundance of interactions in real materials raises the question of direct spectroscopic signatures of phenomena beyond effective single-particle, single-band behaviour. Here we report the identification of quantum oscillations (QOs) in the three-dimensional topological semimetal CoSi, which defy the standard description in two fundamental aspects. First, the oscillation frequency corresponds to the difference of semi-classical quasi-particle (QP) orbits of two bands, which are forbidden as half of the trajectory would oppose the Lorentz force. Second, the oscillations exist up to above 50\,K -- in stark contrast to all other oscillatory components -- which vanish below a few K. Our findings are in excellent agreement with generic model calculations of QOs of the QP lifetime. Since the only precondition for their existence is a non-linear coupling of at least two electronic orbits, e.g., due to QP scattering on defects or collective excitations, such QOs of the QP lifetime are generic for any metal featuring Landau quantization with multiple orbits. They are consistent with certain frequencies in topological semi-metals \cite{2018_vanDelft_PhysRevLett, 2020_Muller_PhysRevResearch, 2021_Ding_JPhysDApplPhys, 2022_Pavlosiuk_PhysRevB, 2023_Broyles_arxiv}, unconventional superconductors\cite{reiss2020quenched, 2015_Sebastian}, rare-earth compounds \cite{2008_McMullan_NJP, 2018_Hiroaki_PRL, 2020_Dalgaard_PRB}, and Rashba-systems \cite{2017_Sunko_Nature}, and permit to identify and gauge correlation phenomena, e.g., in two-dimensional materials \cite{phinney2021strong, 2023_Broyles_PRL} and multiband metals\cite{2019_Sanchez_Nature}.
}


\newpage
\section{Quasiparticle Lifetime in Metals}

In metallic systems with a Fermi surface, Landau quantization \cite{1930_Landau} and the associated QOs \cite{1930_dHvA} represent an invaluable spectroscopic demonstration of the single-particle, independent band character of electronic excitations. Building on the semi-classical approximation of charge carrier motion in which quantum phases are connected to classical trajectories \cite{1984_Shoenberg_Book}, Onsager attributed QOs to the formation of cyclotron orbits with a quantized cross-sectional area. Accordingly, the frequency $f$ of the oscillations is related to the extremal area of an orbit in $k$-space, $A_{\textrm{k}}$, via $f=\hbar/(2\pi e)\,A_{\textrm{k}}$, where $\hbar$ is the reduced Planck constant and $e$ is the electron charge \cite{1933_Peierls_ZPhys, 1952_Onsager_null}. The effects of finite temperatures can be described by an averaging of QOs at zero-temperature with a distribution of different Fermi levels as weighted with a probability given by the negative derivative $-dn_{\textrm{FD}}(E)/dE$ of the Fermi-Dirac distribution $n_{\textrm{FD}}$ \cite{1984_Shoenberg_Book}. As shown by Lifshitz and Kosevich (LK) \cite{1956_Lifschitz_SovPhysJETP}, this results in a damping factor of the amplitude of the oscillations, $R_{\textrm{T}} (m^*)=X/\sinh X$, that depends on the ratio $X=2\pi^2 (k_{\textrm{B}}T)/ (\hbar \omega_c)$ of the thermal energy $k_{\textrm{B}}T$ to the cylcotron energy $\hbar \omega_c=\hbar e B/m^*$. Hence, the thermal damping $R_{\textrm{T}}(m^*)$ is governed by a single-particle effective mass $m^*$ \cite{2004_Kartsovnik_CR, 2015_Julian_book}, where multiband metals are treated as a collection of completely independent bands.

Well-known amendments to the LK-formalism \cite{1984_Shoenberg_Book} comprising quantum tunneling, quantum interference, magnetic interactions, and chemical potential oscillations  maintain a single-particle, independent-band character of the oscillation spectra, whereas recent work on QOs in correlated insulators represents an exception \cite{2015_Tan_Science, 2019_Han_PhysRevLett, knolle2015quantum, sodemann2018quantum}. Likewise, interactions which limit the QP lifetime are so far also treated on a single-particle level of independent bands. Namely, using a Lorentzian distribution of the Landau level energies to reflect the exponential decay associated with the relaxation time approximation, Dingle modelled the decrease of $\tau$ due to interactions in analogy to the LK-formalism \cite{1952_Dingle_ProcRSoL} in terms of a reduction factor $R_{\textrm{D}}=\exp (-\pi m/eB\tau)$ with a scattering time $\tau$ for a given QP band. This energy distribution is often expressed in terms of the Dingle temperature $T_{\rm D} \propto 1/\tau$. 

However, as reported in our paper, the effects of QP scattering are much richer, causing QOs of the QP lifetime (QPL) as a direct consequence of the combination of the Landau quantization of the density of states (DOS) with Fermi's Golden rule. The conditions for QOs of the QPL are remarkably simple: they only require some form of non-linear coupling between FS orbits, i.e., scattering from one FS orbit to the other and back. The resulting QO frequency arises in all physical observables, most prominently electrical transport, but without a corresponding FS cross-section. Thus, QOs of the QPL may be expected in any metal featuring Landau quantization and mutually interacting electronic states. Underscoring this general relevance an analogy exists with so-called magneto-intersubband oscillations (MISOs) in two-dimensional electron gases (2DEGs)\cite{Polyanovsky1988} and quasi-two dimensional metals \cite{Polyanovsky1993}.


\section{Band topology of C\lowercase{o}S\lowercase{i}}

For our study we selected CoSi, a three-dimensional system far from electronic instabilities with a well-defined FS and vanishingly small exchange-enhanced spin splitting. Crystallizing in the chiral space group 198, a comprehensive theoretical assessment recently established a network of topological nodal planes (NP), multifold degeneracies, and Weyl points \cite{2022_Huber_PhysRevLett} (Methods). Shown in Fig.~\ref{fig:2}\,a is the calculated electronic structure taking into account spin-orbit coupling, where the presence of NPs (green shading) on the surface of the cubic Brillouin zone (BZ), as well as a sixfold degenerate point at R are highlighted \cite{2019_Rao_Nature, 2019_Sanchez_Nature}.  Key for our study are the strikingly simple QO spectra associated with the R point of the BZ \cite{2022_Huber_PhysRevLett, 2022_Guo_NatPhys}. Whereas four almost parallel bands originate from the sixfold degeneracy at R (Fig.~\ref{fig:2}\,b), thus forming four small intersecting electron pockets (Fig.~\ref{fig:2}\,c), only two dominant frequencies characteristic of two FS orbits are observed for all field orientations. This may be traced to exact degeneracies enforced by the nodal planes at the R point and additional quasi-degeneracies, being well-described in terms of two adjacent FS orbits.

For instance, for $B\parallel [001]$, the dispersion in the plane of the extremal orbits (green) corresponds to two wine goblet shapes that are twofold degenerate everywhere (left hand side of Fig.~\ref{fig:2}\,d). The associated extremal orbits at the FS are here given by two concentric rings (grey lines on the right hand side of Fig.~\ref{fig:2}\,d). The same is effectively true for $B$ away from $B\parallel [001]$, as illustrated for $B\parallel [111]$ and $B\parallel [110]$ in Fig.~\ref{fig:2}\,e and \ref{fig:2}\,f, respectively.  While the FS contours (coloured lines) seem more complicated, the degeneracies enforced by the nodal planes and the additional quasi-degeneracies result in four orbits with pairwise identical cross sections. In a recent study, it has been proposed that the QO spectra at the R point are due to hidden quasi-symmetries \cite{2022_Guo_NatPhys}, resulting in the same four orbits with pairwise identical cross sections (see also Extended Data Fig.\,\ref{fig:EDI3}).

\section{Shubnikov--de Haas oscillations}

\begin{figure*}
	\centering\includegraphics[width=0.45\textwidth,clip=]{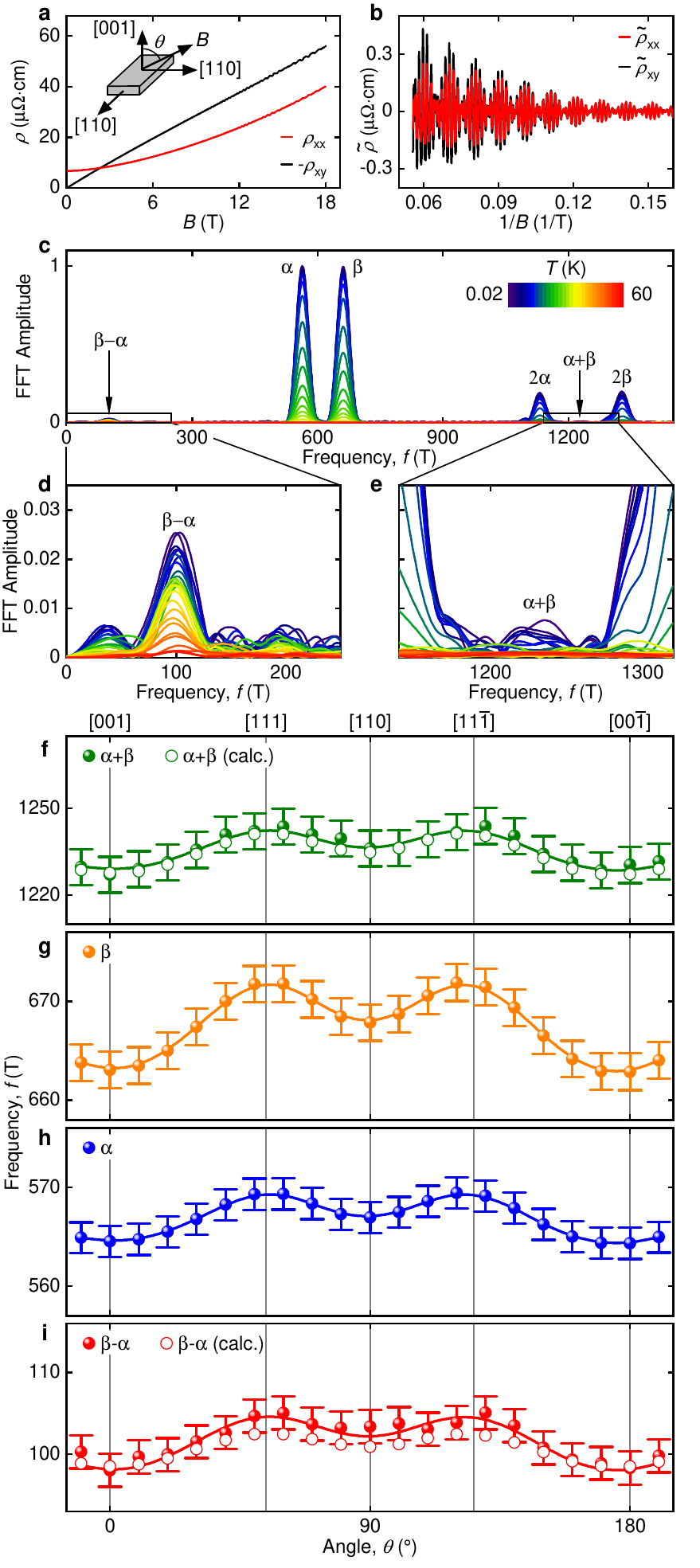}
\linespread{1.0}\selectfont{}
\caption{\raggedright
{\bf $|$ Quantum oscillation spectrum of CoSi associated with the R point.}
{\bf a},~Transverse magnetoresistivity, \rxx{}, and Hall resistivity, \ryx{}, as a function of magnetic field, $B$, at $T=20$\,mK and $\theta=0^{\circ}$. The inset shows the experimental geometry, where $B$ is rotated in the (110) plane and $\theta$ denotes the angle between [001] and $B$. {\bf b},~Oscillatory signal components of \rxx{} and \ryx{} after subtraction of a smooth non-oscillatory background as a function of inverse magnetic field. {\bf c},~Typical FFT spectra of the data shown in b in the temperature range between $T=20$\,mK and 60\,K. {\bf d-e},~Close-up views of the frequency regime of the difference and the sum of the fundamental frequencies $\alpha$ and $\beta$. {\bf f-i},~Oscillation frequencies as a function of $\theta$ at $T=300$\,mK.  Open circles denote the calculated sum and difference of the frequencies $f_\alpha$ and $f_\beta$. The angular dependence further establishes the identification of the detected frequencies $f_{\alpha+\beta}$ and $f_{\beta-\alpha}$ as combinations of the fundamental frequencies $f_\alpha$ and $f_\beta$.
}
\label{fig:3}
\end{figure*}

Shown in Fig.~\ref{fig:3}\,a are the transverse magnetoresistance, \rxx{}, and the Hall resistivity, \ryx{}, at 20\,mK (Methods). Both, \rxx{} and \ryx{}, exhibit pronounced oscillations for magnetic fields exceeding $\sim 6\,\mathrm{T}$. Subtracting a monotonic background, the oscillatory signal components, \tilrxx{} and \tilryx{}, were found to display the same QO spectra (Fig.~\ref{fig:3}\,b and Methods). Periodic in $1/B$, a pronounced beating pattern characteristic of two nearly identical frequencies dominates the data. Typical fast Fourier transforms (FFTs) for magnetic field parallel to $[001]$ are shown in Fig.~\ref{fig:3}\,c. Two prominent peaks at $f_{\alpha}=565\,\mathrm{T}$ and $f_{\beta}=663\,\mathrm{T}$ with up to three higher harmonics may be resolved due to FS sheets centred on the R point\cite{2022_Huber_PhysRevLett, 2022_Guo_NatPhys} (Methods). In the following we focus on \tilrxx{}; \tilryx{} is presented in Extended Data Fig.\,\ref{fig:EDI1}.

Not reported before are the two additional frequencies shown in Figs.~\ref{fig:3}\,d and \ref{fig:3}\,e, which correspond to the difference, $f_{\beta - \alpha}$, and the sum, $f_{\beta+ \alpha}$, of the fundamental frequencies $f_{\alpha}$ and $f_{\beta}$. As a function of field direction a small but unambiguous angle dependence may be discerned for $f_{\beta+ \alpha}$, $f_{\beta}$, $f_{\alpha}$, and $f_{\beta - \alpha}$ as shown in Fig.~\ref{fig:3}\,f through i, where the variations of $f_{\beta}$ and $f_{\alpha}$ agree with the Fermi surface calculated in density function theory (DFT) \cite{2022_Huber_PhysRevLett}. Importantly, the magnitude of the variations of $f_{\beta}$ and $f_{\alpha}$ differs quantitatively by almost a factor of two amounting to $9$ and $5\,\mathrm{T}$, respectively. This implies a distinct quantitative difference of the angle dependence of the calculated sum and difference of $f_{\beta}$ and $f_{\alpha}$ in excellent agreement with experiment, and shows that $f_{\beta - \alpha}$ and $f_{\beta+ \alpha}$ originate in the difference and the sum of $f_{\alpha}$ and $f_{\beta}$ (cf. Extended Data Fig.\,\ref{fig:EDI4} and methods).

The magnetic field dependence of $f_\alpha$ and $f_\beta$ corresponds to Dingle temperatures of $T_{\rm D\alpha, th}\sim1.3\,{\rm K}$ and $T_{\rm D\beta, th}\sim1.2\,{\rm K}$. The associated quasiparticle lifetimes are of the order $\tau_{\rm QP}\sim10^{-12}\,{\rm s}$ with a Fermi velocity $v_{\rm F}\sim 3.4*10^5\,{\rm m/s}$, consistent with literature \cite{2022_Guo_NatPhys}. In turn, the mean free path is $l\sim3.4*10^{-7}\,{\rm m}$, characteristic of a tiny defect concentration. Unfortunately, independent microscopic determination of the precise nature and concentration of  defects at these low levels is not possible, posing a major challenge in condensed matter physics on the whole. 

\section{Temperature dependence}

\begin{figure}[ht]
	\centerline{\includegraphics[width=0.45\textwidth,clip=]{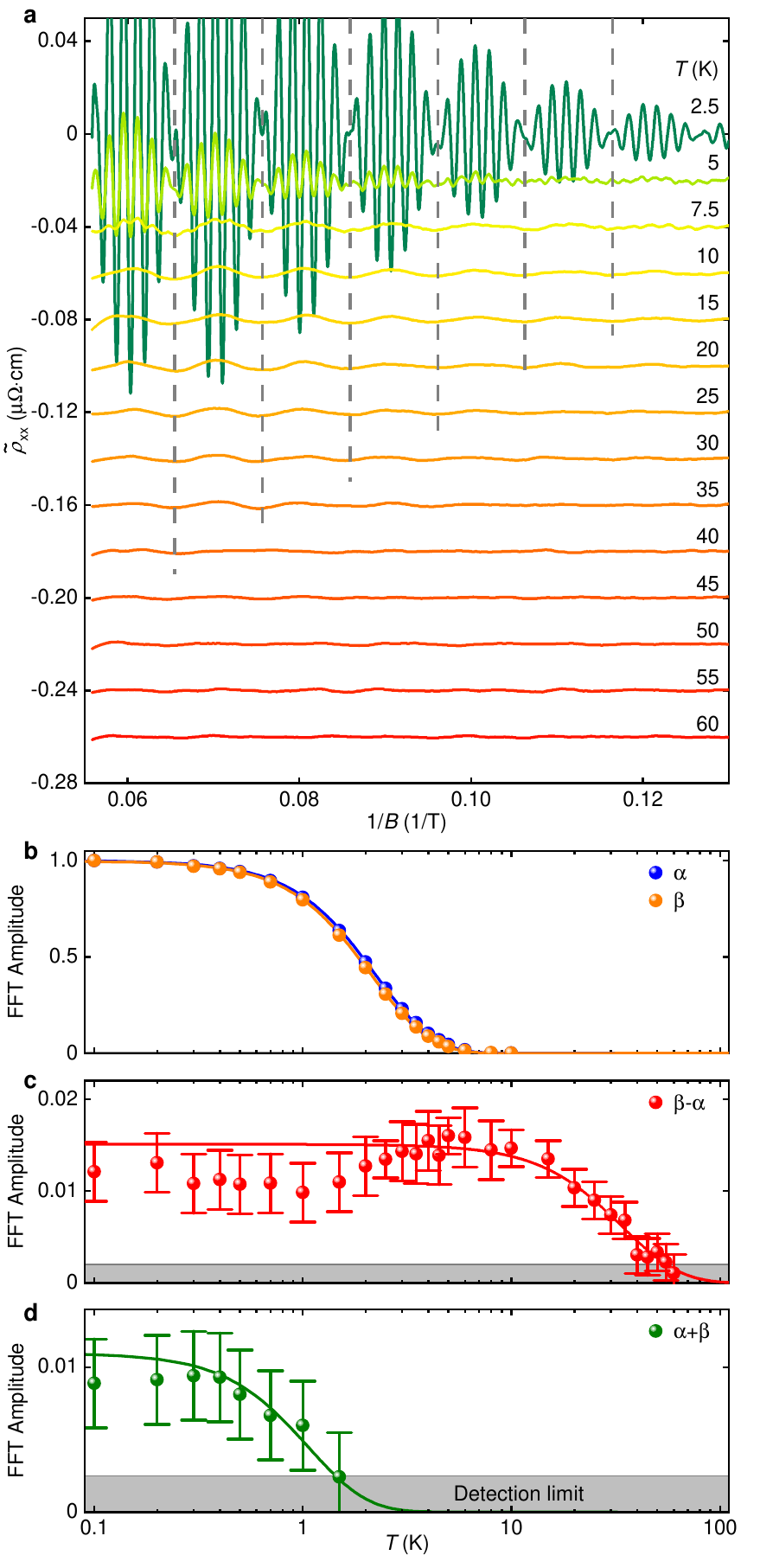}}
\linespread{1.0}\selectfont{}
\caption{\raggedright
{\bf $|$ Temperature dependence of Shubnikov--de~Haas oscillations in the transverse magnetoresistance of CoSi.}
Oscillation amplitudes shown in panels b-d were analyzed in the magnetic field range between 9 and 18\,T with $B$ applied along the [001] direction and normalized to the amplitude of $f_{\alpha}$ at $T=20$\,mK. 
{\bf a}, Oscillatory component of \rxx{} as a function of inverse magnetic field at different temperatures. Curves are shifted for better visibility.
{\bf b},~Oscillation amplitudes of $f_{\alpha}$ and $f_{\beta}$ as a function of temperature $T$. Lines represent a fit of the temperature reduction factor $R_T$ in the LK formalism. The inferred effective masses are $m_{\alpha}^{*}=(0.92 \pm 0.01)\,m_\mathrm{e}$ and $m_{\beta}^{*}=(0.96 \pm 0.01)\,m_\mathrm{e}$, in good agreement with the effective masses obtained from \ryx{} (Extended Data Fig.\,\ref{fig:EDI1}). Error bars are smaller than the data points.
{\bf c},~Oscillation amplitudes of the difference frequency $f_{\beta-\alpha}$. A fit of the data between $2$\,K and $60$\,K with the LK temperature reduction factor $R_{\rm T}$ is represented by the red line and yields an effective mass of $m_{\beta-\alpha}^{*}=(0.06 \pm 0.01)\,m_\mathrm{e}$. 
{\bf d},~Oscillation amplitudes of frequency $f_{\alpha+\beta}$. The line represents a LK fit yielding an effective mass of $m_{\alpha+\beta}^{*}=(1.9 \pm 0.4)\,m_\mathrm{e}$. 
}
\label{fig:4}
\end{figure}

As shown in  Fig.~\ref{fig:4}\,a, with increasing temperature the oscillations at $f_{\alpha}$ and $f_{\beta}$ vanish, while the oscillations at $f_{\beta-\alpha}$ may be discerned up to at least 50\,K. The agreement between the beating pattern in $f_{\alpha}$ and $f_{\beta}$ at low temperatures with the phase of the oscillations at $f_{\beta-\alpha}$ at elevated temperatures (cf Extended Data Figs.\,\ref{fig:EDI1} and \ref{fig:EDI2}), corroborates that $f_{\beta-\alpha}$ and $f_{\alpha+\beta}$ originate in $f_{\alpha}$ and $f_{\beta}$. Shown in Fig.~\ref{fig:4}\,b through \ref{fig:4}\,d is the temperature dependence of the oscillation amplitude at $f_{\alpha}$, $f_{\beta}$, $f_{\beta-\alpha}$, and $f_{\alpha+\beta}$ as normalized to the amplitude of $f_{\alpha}$ for $T\to0$. The small size of the amplitudes at $f_{\beta-\alpha}$ and $f_{\alpha+\beta}$ underscores the high signal to noise ratio of our experimental set-up. 

The temperature dependence of $f_{\alpha}$ and $f_{\beta}$ follows accurately the standard LK-formalism, where we find cyclotron masses of $m^*_\alpha = (0.92 \pm 0.01)~m_\mathrm{e}$ and $m^*_\beta = (0.96 \pm 0.01)~m_\mathrm{e}$, respectively, consistent with previous reports \cite{2022_Huber_PhysRevLett, 2022_Guo_NatPhys} (Methods). The presence of $f_{\beta-\alpha}$ up to at least $\sim$50\,K shown in Fig.~\ref{fig:4}\,c is highly unusual. Fitting the temperature dependence between 2 and 60\,K yields an effective mass $m^*_{\beta-\alpha}=(0.06\pm 0.01)\,m_\mathrm{e}$ consistent with $m^*_\beta - m^*_\alpha$. The amplitude of $f_{\alpha+\beta}$, shown in Fig.~\ref{fig:4}\,d is, finally, consistent with $m^*_{\alpha+\beta}=(1.9\pm 0.3)\,m_\mathrm{e}$.

In comparison to \tilrxx{} the oscillations at $f_{\beta-\alpha}$ observed in \tilryx{}, shown in Extended Data Fig.\,\ref{fig:EDI1}, are characteristic of $m^*_{\rm HT} = (0.07\pm 0.01)\,m_\mathrm{e}$ (denoted HT). Interestingly, the temperature dependence suggests an additional contribution at low temperatures (denoted LT) consistent either with a decay $R_{\textrm{T}}(m^*_\alpha) R_{\textrm{T}}(m^*_\beta)$ or with $R_{\textrm{T}}(m^*_{\rm LT})$ and $m^*_{\rm LT} = (1.6\pm 0.3)\,m_\mathrm{e} \approx m^*_\beta + m^*_\alpha$.

\section{Theoretical model}

Our main new findings to be explained theoretically are as follows. First, we observe oscillations at $f_{\beta-\alpha}=f_{\beta}-f_{\alpha}$, and $f_{\alpha+\beta}=f_{\alpha}+f_{\beta}$, of the frequencies $f_{\alpha}$ and $f_{\beta}$. Second, as a function of field orientation tiny variations of $f_{\alpha+\beta}$ and $f_{\beta-\alpha}$ closely track the values calculated from $f_{\alpha}$ and $f_{\beta}$. Third, as a function of temperature the FFT amplitudes of $f_{\alpha}$ and $f_{\beta}$ decrease characteristic of nearly identical masses $m_{\alpha}^{*}=(0.92 \pm 0.01)\,m_\mathrm{e}$ and $m_{\beta}^{*}=(0.96 \pm 0.01)\,m_\mathrm{e}$, respectively. Fourth, as a function of temperature the FFT amplitude of $f_{\beta-\alpha}$ survives up to unusually high temperatures, in agreement with $m_{\beta}^{*}-m_{\alpha}^{*}$, while the FFT amplitude of $f_{\alpha+\beta}$ vanishes at much lower temperatures consistent with $m_{\beta}^{*}+m_{\alpha}^{*}$. Fifth, the phase of the beating pattern of $f_{\alpha}$ and $f_{\beta}$ at low temperatures is fixed with respect to the phase of $f_{\beta-\alpha}$ at high temperatures. 

The properties of CoSi rule out conventional mechanisms as an account of the new oscillation frequencies (see Methods for details) \cite{1984_Shoenberg_Book, 2007_Alexandrov_PhysRevB, 2021_Allocca_arXiv}. Namely, for magnetic breakdown (MB) \cite{1961_Cohen_PRL, 1962_Blount_PhysRev, 1966_Chambers_ProcPhysSoc} the electron-like FS orbits would require a motion against the Lorentz force for half of the trajectory associated with $f_{\beta}-f_{\alpha}$, the amplitude of which would be suppressed much faster with increasing temperature. Chemical potential oscillations are well known in two-dimensional systems, but will be quantitatively irrelevant for the nearly isotropic three-dimensional FS sheets of CoSi. For magnetic interactions the values of $d\tilde{M}/dH$ are orders of magnitude too small to account for our experiment (cf Extended Data Fig.\,\ref{fig:EDI6}). Quantum interference, finally, would require a superposition between different pathways connected by MB junctions\cite{1984_Bergmann, 1985_Lee-UCFs}, which do not exist for $B\parallel [001]$ (Fig.\,\ref{fig:2}\,d) and do not have suitable MB probabilities for other field directions (Extended Data Fig.\,\ref{fig:EDI3}).

\begin{figure}[ht]
	\centerline{\includegraphics[width=1.0\textwidth,clip=]{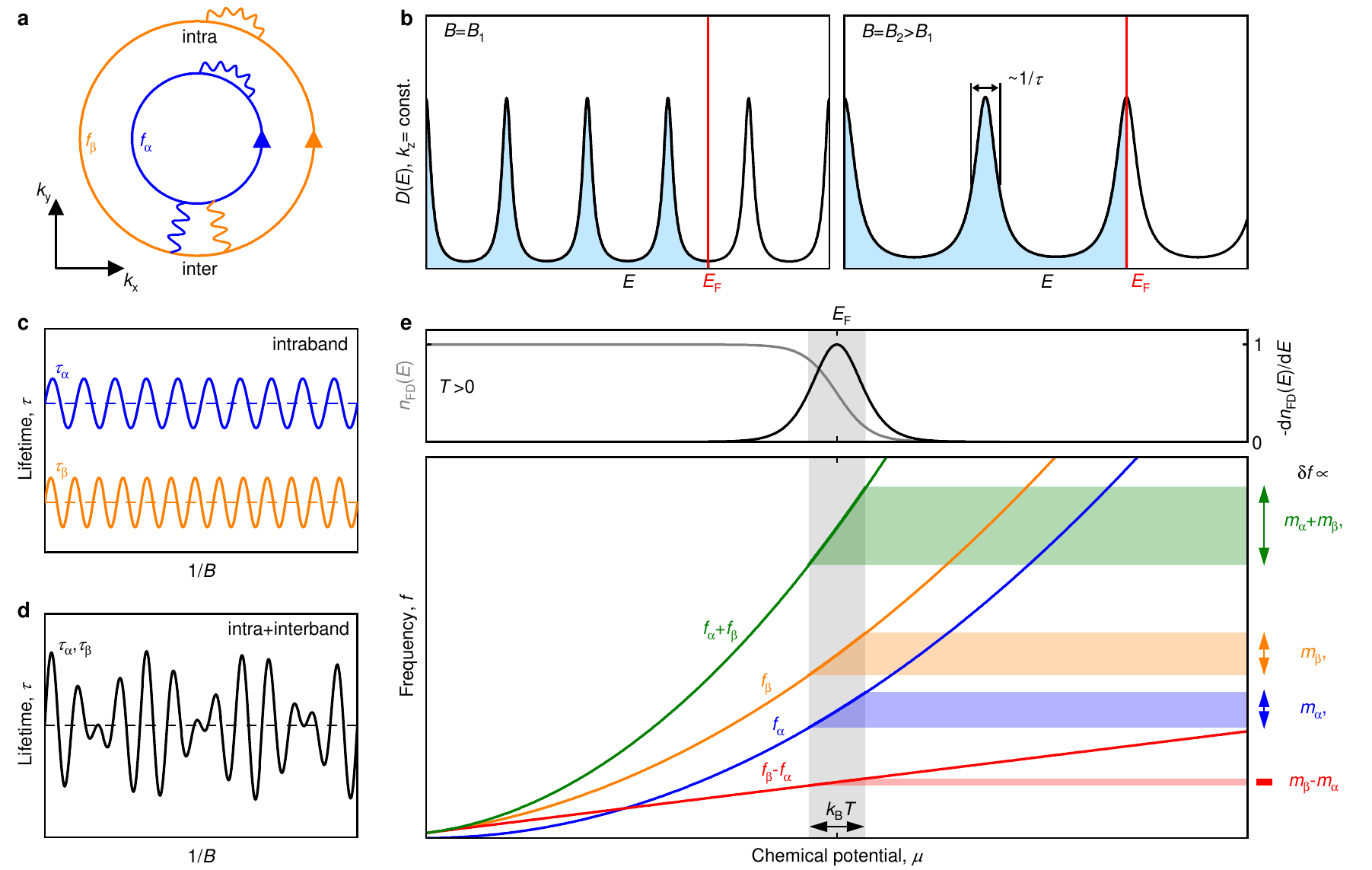}}
\linespread{1.0}\selectfont{}
\caption{\raggedright
{\bf $|$ Origin of quantum oscillations of the quasiparticle lifetime.} 
{\bf a}, Depiction of two Fermi surface orbits arising from nested bands. Associated with the two FS orbits are two oscillation frequencies $f_\alpha$ and $f_\beta$. Intraband scattering causes transitions within each FS orbit; interband scattering causes transitions between the FS orbits. The scattering effectively mixes the orbits. For the associated Feynman diagrams see Methods. {\bf b}, Density of states, $D(E)$, due to Landau quantization of cyclotron orbits under magnetic fields $B_1$ and $B_2>B_1$. The Landau level separation and degeneracy varies linearly with B such that the density of states at the Fermi energy $E_{\rm F}$ is an oscillatory function of $1/B$. Intraband transitions cause a broadening of the Landau levels. As the lifetime $\tau_1$ of the cyclotron orbits varies with $D(E_{\rm F})$, it is also an oscillatory function of $1/B$. {\bf c}, Quasiparticle lifetime associated with FS orbits $\alpha$ and $\beta$ with intra-orbit but no inter-orbit scattering.  {\bf d}, Quasiparticle lifetime associated with FS orbit $\alpha$ and $\beta$ in the presence of intra-orbit \textit{and} inter-orbit scattering. As the inter-orbit scattering depends on the density of states on both FS orbits $\alpha$ and $\beta$, the QPL oscillates with a superposition of the densities of states on both orbits. {\bf e}, Connection between the temperature dependence of the oscillation amplitude and the effective mass as proposed by Lifshitz and Kosevitch. The nearly parallel dispersion of the orbits $\alpha$ and $\beta$ as a function of energy near the chemical potential results in a nearly complete suppression (doubling) of the dispersion for oscillations at $f_{\beta-\alpha}$ ($f_{\alpha+\beta}$). The temperature dependence is governed by the variation of frequency $\delta f$ in an interval $k_{\rm B}T$ around $E_{\rm F}$ depicted in grey shading given by the negative of the derivative of the Fermi distribution with respect to energy, $-d n_{\rm FD}/dE$.
}
\label{fig:1}
\end{figure}

Instead, our findings can be explained in terms of QOs of the QPL, as summarized in Fig.\,\ref{fig:1}, where the single precondition is some form of non-linear coupling that generates  intra-orbit \textit{and} inter-orbit transitions as illustrated in Fig.\,\ref{fig:1}\,a. To demonstrate the plausibility of the non-linear coupling for a specific microscopic situation, we considered defect scattering (see Methods). A well-known consequence of QP transitions is a broadening of the Landau levels as depicted in Fig.\,\ref{fig:1}\,b, where the inverse of the QP lifetime $1/\tau$ corresponds to the half-width of the DOS peaks. Assuming the Born approximation and Fermi's Golden rule, the QPL varies with the DOS at $E_{\rm F}$. Hence, as a function of magnetic field $\tau$ includes an oscillatory component in addition to the average value. 

In the presence of intra-orbit scattering, the oscillatory component of $\tau$ varies with the cyclotron frequency of the underlying FS orbit as shown in Fig.\,\ref{fig:1}\,c. Remarkably, in the presence of intra-orbit \textit{and} inter-orbit transitions, $\tau$ oscillates with both frequencies of the participating FS orbits, $f_\alpha$ and $f_\beta$, as depicted in Fig.\,\ref{fig:1}\,d where the intra-orbit \textit{and} inter-orbit oscillations were assumed to be equally strong. Expressed in terms of the Dingle temperature associated with the QP lifetime,  $T_{\rm D}$, and the Dingle temperature of the non-oscillatory average, $T^0_{\rm D}$, the inverse of the QPL may be written as 
\begin{equation}
\frac{1}{\tau} \propto T_{\rm D} \propto T^0_{\rm D} + A_{\alpha}\cos( \frac{2\pi f_{\alpha}}{B})+ A_{\beta}\cos( \frac{2\pi f_{\beta}}{B})
\label{eq:simple_tau}
\end{equation}
where $A_{\alpha}$ and $A_{\beta}$ are prefactors, and $B$ is the magnetic field.

In the physical quantities studied experimentally, the QOs of the QPL cause additional oscillatory components expressed conveniently in terms of $T_{\rm D}$. Due to an additional linear dependence on $T_{\rm D}$, the effect is particularly pronounced in the electrical transport properties
\begin{equation}
\sigma \propto T_{\rm D} \left[\cos( \frac{2\pi f_{\alpha}}{B})+\cos( \frac{2\pi f_{\beta}}{B}) \right] \,R_{\rm D} (T_{\rm D}).
\label{eq:simple_cond}
\end{equation}
For the sake of the argument, it is not necessary to distinguish between \tilrxx{} and \tilryx{} here. Inserting $T_{\rm D}$, one obtains to leading order terms in \mbox{$\cos\left( 2\pi f_{\alpha}/B\right)\times\cos( 2\pi f_{\beta}/B)$} and $\cos^2( 2\pi f_{\lambda}/B)$ ($\lambda=\alpha,\beta$). Using trigonometric addition theorems, the former decomposes into $\cos(2\pi (f_{\beta}-f_{\alpha})/B)$ and $\cos(2\pi (f_{\beta}+f_{\alpha})/B)$. The combined effect of intra-orbit and inter-orbit transitions, hence, accounts naturally for the nearly isotropic oscillatory components at $f_\beta-f_\alpha$ and $f_\alpha+f_\beta$ observed experimentally. Moreover, it explains naturally the fixed phase relationship of $f_\beta-f_\alpha$ with $f_\alpha$ and $f_\beta$ observed experimentally (Extended Data Fig.\,\ref{fig:EDI2}). Further subleading oscillatory dependencies we ignore for clarity, such as $2 f_\alpha-f_\beta$, $2 f_\beta -f_\alpha$, $2 f_\beta- 2 f_\alpha$, $3 f_\beta - f_\alpha$, $2 f_\alpha$, $2 f_\beta$, $3 f_\alpha$ and so forth, arise from the Dingle damping factor itself, $R_{\rm D} (T_{\rm D})$ (see Methods). As for the terms in $\cos^2( 2\pi f_{\lambda}/B)$ ($\lambda=1,2$), which are purely due to intra-orbit transitions, these will only modify the higher harmonics of the fundamental frequencies. 

It is now instructive to address the effects of finite temperatures as illustrated in Fig.\,\ref{fig:1}\,e. The fundamental QO frequencies are determined by the extremal cross sectional areas $A_{\textrm{k}}(E_{\mathrm{F}})$. Accordingly, oscillations at $f_\beta-f_\alpha$ and $f_\alpha+f_\beta$ also change as a function of  $E_{\mathrm{F}}$. Recalling that $R_{\rm T}(T)$ originates from an averaging of orbits with different $E_{\mathrm{F}}$, weighted by $-dn_{\mathrm{FD}}/dE$, one finds that the width of the distribution of Fermi levels at finite temperature, $\delta E_{\textrm{F}}\sim k_{\mathrm{B}}T$, causes a distribution of frequencies, $\delta f$, that scales with the cyclotron mass, $\delta f/k_{\mathrm{B}}T=\hbar/(2\pi e)\, \partial A_{\textrm{k}}/\partial E = m/(e\hbar)$. Consequently, for the oscillations at $f_\beta-f_\alpha$ and $f_\alpha+f_\beta$ finite temperatures cause distributions given by $\delta (f_\beta-f_\alpha)/k_{\mathrm{B}}T=\hbar/(2\pi e) \,\partial (A_{\textrm{k},\beta}-A_{\textrm{k},\alpha})/\partial E =(m_\beta-m_\alpha)/(e\hbar)$ and $\delta (f_\alpha+f_\beta)/k_{\mathrm{B}}T=(m_\beta+m_\alpha)/(e\hbar)$, respectively. Taken together, this explains the unusual temperature dependence of the oscillation amplitudes $f_\beta-f_\alpha$ and $f_\alpha+f_\beta$ as compared with $f_\alpha$ and $f_\beta$.

QOs of the QPL are generic and may be expected for any FS orbits coupled non-linearily regardless of the coupling mechanism and the specific orbits involved. Thus, they may be intrinsic in perfectly pure systems, arising, e.g., from interactions with collective excitations. In case the non-linear coupling is due to defects, a sweet spot of the concentration is expected at which intra-orbit scattering is sufficiently weak such that Landau quantization persists, while inter-orbit scattering is sufficiently strong to couple the bands. In real materials the same microscopic defects give rise to both intra- and inter-orbit transitions.

It is also instructive to consider the implications of different band dispersions (see also Extended Data Fig.\,\ref{fig:EDI7}). For instance, if the cyclotron masses of $f_\alpha$ and $f_\beta$ are equal over an energy range of order $k_{\mathrm{B}}T$ around $E_{\textrm{F}}$ regardless of the full dispersion relation, a complete suppression of the temperature decay is expected. Such a behaviour may be erroneously taken as evidence of  fermions with vanishing cyclotron mass. A possible explanation for the additional enhancement of the oscillation amplitude of $f_{\beta-\alpha}$ below $\sim$1\,K, seen in \tilryx{}, may be due to other coupling mechanisms like Coulomb repulsion, as recently discussed for quasi-2D systems \cite{2021_Allocca_arXiv}. Importantly, all of these mechanisms predict a strong temperature dependence  \cite{2021_Allocca_arXiv,1984_Shoenberg_Book}, i.e., $R_{\rm T}(m_\alpha^*)R_{\rm T}(m_\beta^*)$, which is inherently different to our theory.

As stated above, QOs of the QPL are particularly pronounced in the transport properties due to the additional dependence on the QPL. In turn, similar QO frequencies albeit with a different amplitude and temperature dependence are also expected in the magnetization, notably the de Haas-van Alphen effect, as well as all other physical quantities. Thus, differences of spectral weight of QOs observed in different physical quantities will permit to identify QOs of the QPL.

\section{Discussion}

QO of the QPL are generic for any FS orbits coupled non-linearily. They may be identified most easily for the two electron-like FS sheets in CoSi. Such interband-QO of the QPL may also be expected for electron- and hole-like FS sheets. Indeed, a sum frequency between an electron- and a hole-like FS sheet has been reported in Co$_3$Sn$_2$S$_2$, HfSiS, and UTe$_2$, where they would resolve the inconsistencies with conventional mechanisms as noted in these papers \cite{2018_vanDelft_PhysRevLett, 2021_Ding_JPhysDApplPhys, 2023_Broyles_arxiv}. Similarly, QO of the QPL may also arise in anisotropic single-band systems between orbits on different extremal cross-sections on the same FS sheet. Such intraband-QOs of the QPL may have been observed unknowingly in MoSi$_2$ and WSi$_2$ \cite{2022_Pavlosiuk_PhysRevB}, where disagreement of a difference frequency with conventional mechanisms was emphasized. Further, efforts to identify QO frequencies that are not present in DFT has motivated proposals of FS reconstructions. However, QO of the QPL may allow to reconcile some QO spectra without need for FS reconstructions. Examples may exist in the FeAs- or kagome-superconductors \cite{reiss2020quenched, fu2021quantum, 2023_Broyles_PRL}. Indeed, one might even reassess the cuprate superconductors, representing single-band systems\cite{2015_Sebastian}. Similarly, in heavy-fermion compounds unexplained QOs with moderately heavy masses may be due to FS orbits with very heavy masses that evade detection down to the low mK-regime \cite{1987_Taillefer_JMMM, 2008_McMullan_NJP, 2018_Hiroaki_PRL, 2020_Dalgaard_PRB}. 

QOs of the QPL may be also of great relevance in spin-orbit coupled systems with Rashba- or Dresselhaus-type dispersions, or -- in general -- for nested FS pockets that enclose (near) degeneracies of two or more bands. In bulk multi-valley systems QOs of the QPL may arise from FS pockets not centered around a common momentum vector in the presence of large-momentum transitions, e.g., mediated by spin-, charge-,  or phonon-fluctuations \cite{2023_Allocca_arXiv}. This will help to construct more accurate models of quantum materials and may lead to applications in, e.g., valleytronics. Taken together, QO of the QPL quite generally allow to quantify the strength of QP scattering in bulk and tailored materials as reported recently in twisted bilayer graphene  \cite{phinney2021strong, 2023_Broyles_PRL}.


\newpage
\section*{References}


\newpage
\section*{Main figures}

\clearpage \thispagestyle{empty}

\captionsetup[figure]{labelfont={bf},name={Fig.},labelsep=space}

\clearpage \thispagestyle{empty}

\begin{figure*}[ht]
	\centerline{\includegraphics[width=1.0\textwidth,clip=]{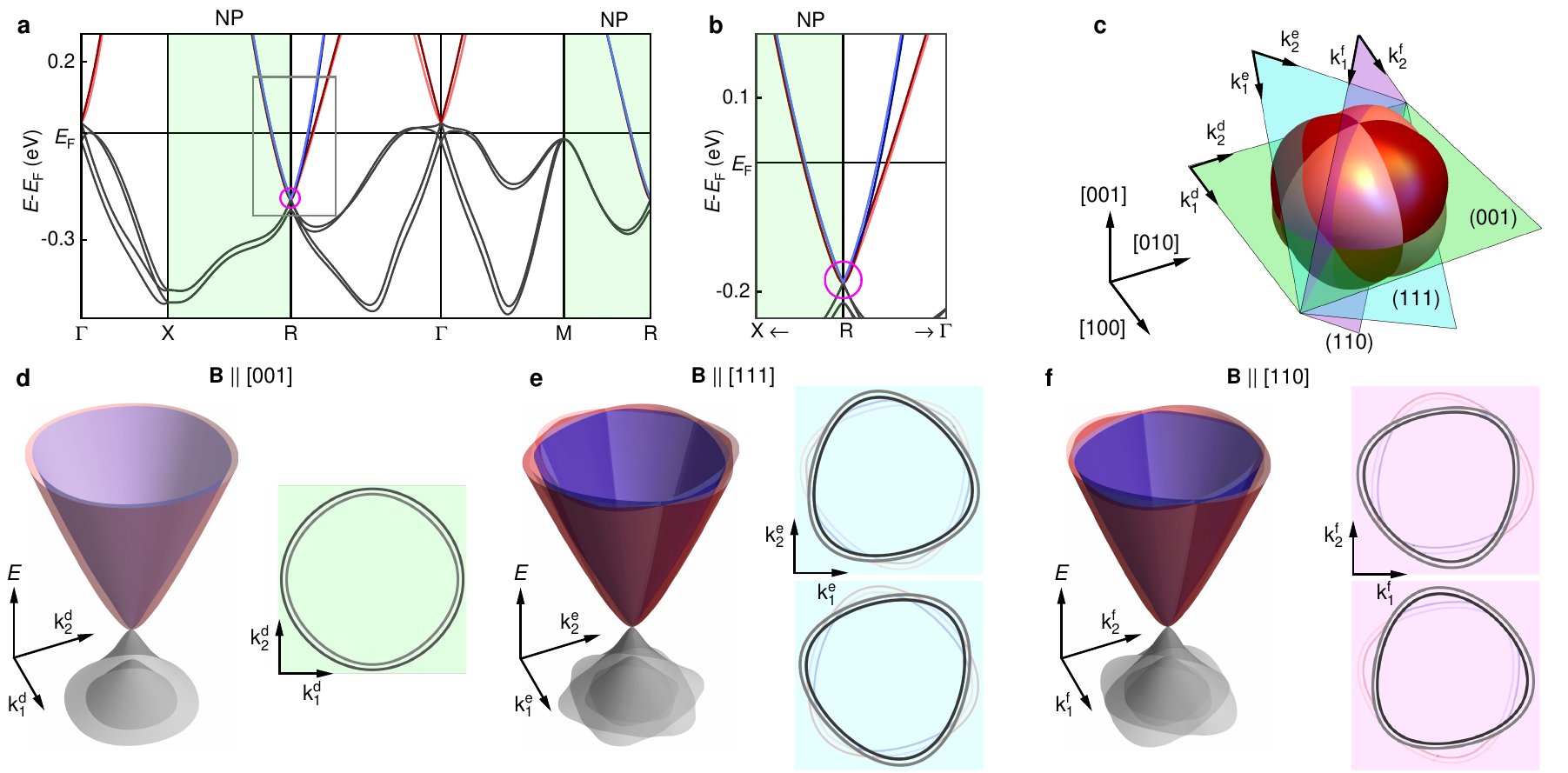}}
\linespread{1.0}\selectfont{}
\caption{\raggedright
{\bf $|$ Electronic band structure of CoSi and Fermi surface orbits featuring quantum oscillations of the quasiparticle lifetime.} For any field direction, only two different extremal FS cross-sectional areas exist around the R point. Topological nodal planes (NPs) are shown in green shading. See also Extended Data Fig.\,\ref{fig:EDI3} for details why effectively two FS orbits dominate the spectra.
{\bf a}, Electronic band structure of CoSi. {\bf b}, Multi-fold band crossings and nested dispersion near $E_{\rm F}$ in the vicinity of the intersection of three topological nodal planes at the R point. {\bf c}, Fermi surface pockets around the R point. Four nearly spherical FS sheets intersect at the nodal planes. Planes of extremal orbits for selected field directions as shown in panels d through f are denoted by different color-shading. Directions in momentum space spanning these planes are denoted with a superscript referring to panels d, e, and f. {\bf d}, Dispersion for $B\parallel [001]$ (left) and FS orbits as seen in the direction of the applied field (right, grey lines). The FS orbits reside within a nodal plane.  {\bf e}, Dispersion for $B\parallel [111]$ (left) and FS orbits as seen in the direction of the applied field (right, colored lines). {\bf f}, As in panels (d) and (e) for $B\parallel [110]$. In all cases, the cross-sectional areas are pairwise degenerate.
}
\label{fig:2}
\end{figure*}

\clearpage \thispagestyle{empty}

\clearpage \thispagestyle{empty}

\clearpage \thispagestyle{empty}


\clearpage
\newpage
\section*{Methods}

\subsection{Experimental methods and data analysis}

Several studies have been reported in the literature that addressed the electronic structure of CoSi\cite{2022_Huber_PhysRevLett, 2019_Rao_Nature, 2022_Guo_NatPhys, 2019_Yuan_ScienceAdvances, 2019_Wu_ChinesePhysLett, 2019_Xu_PhysRevB, 2020_Wang_PhysRevB, Sasmal2022}. Our experimental results are in excellent agreement with the literature, where comparison is possible.

\textbf{Crystal growth and sample preparation}:
Several single crystals of CoSi were prepared either by optical float zoning \cite{Neubauer-RSI-2011, Bauer-RSI-2016} or flux growth using Te as flux. The sample investigated most extensively of which data are reported in this manuscript, was grown by optical float zoning. The ingot was oriented by x-ray Laue diffraction. Using a wire saw a platelet was cut, with dimensions 2.1x1.0x0.2\,mm$^3$ with faces perpendicular to $[110]$, $[110]$, and $[001]$, respectively. The platelet was subsequently annealed under Ar atmosphere at $1100$\,$^{\circ}$C for 100\,h to improve the sample purity. High sample quality was confirmed by means of the electrical resistivity, magnetization, specific heat and x-ray scattering. The residual resistivity ratio of this sample was 16. Selected measurements of further samples featuring residual resistivity ratios up to 32 confirmed all of the observations reported here. No dependence of the QO of the QPL on the relatively small variations of the residual resistivity ratios in our samples was observed.\\

\textbf{Shubnikov-de Haas measurements}:
The electrical transport measurements were performed using a $^3$He-$^4$He dilution refrigerator equipped with an $18\,{\rm T}$ ($20\,{\rm T}$) superconducting magnet system. For measurements above $\sim 10\,{\rm K}$ a variable temperature insert was used, where data agreed in the overlapping temperature range down to 1.5\,K. The sample was contacted electrically by means of wire bonds arranged in a standard six-point Hall configuration. The longitudinal voltage contacts were placed near the edge of the sample. Vanishingly small asymmetries of the longitudinal and transverse voltage pick-up under field-reversal confirmed essentially perfect contact geometry. Electrical excitation currents at a frequency of 22.08\,Hz were used, where the absence of Joule heating was confirmed using different excitation amplitudes. The longitudinal and transverse voltage pick-up were amplified with impedance-matching transformers selected to match the pre-amplifiers used. The voltage pick-ups were recorded simultaneously using digital lock-in detectors. In the dilution refrigerator the impedance-matching transformer was operated at 1.5\,K, reducing the effect of Johnson noise. A piezoelectric rotation stage was used to change the sample orientation in-situ. The electrical current was applied along the [110] crystallographic direction and the sample was rotated in the (110) plane, perpendicular to the current direction. 

For a pedagogical step-by-step illustration of the procedure used to analyze the data we refer to the supplement of Ref.\,\onlinecite{2022_Huber_PhysRevLett}. The non-oscillatory signal contributions in  \ryx{} and \rxx{} were approximated by polynomial fits of 5$^{th}$ and 6$^{th}$ order, respectively, and subsequently subtracted. Remaining contributions that were not periodic in $1/B$ due to imperfect background subtraction were removed, subtracting a moving average. The quantum oscillation spectra were confirmed to be insensitive to the precise choice of field range analysed, where data was recorded up to 20\,T for selected temperatures and sample orientations. Spectra shown in our manuscript  were analysed for fields between $9$ and $18$\,T. For the Fourier analysis the data was interpolated at equidistant points in $1/B$. To avoid oversampling, the spacing of the points in $1/B$ corresponded to the spacing at which data was recorded at the highest fields. Between 0 and $1\,{\rm T^{-1}}$ additional zeros were added on either side of the range recorded experimentally in order to enhance the sampling rate of the FFT algorithm (this is known as zero padding). To reduce spectral leakage a Hamming window was applied before calculating the FFT. 

Peaks of the FFT spectra were fitted with Gaussians to determine the oscillation frequencies and amplitudes. Both up- and downsweep of the magnetic field were recorded for each temperature and field orientation and the mean of the oscillation amplitudes and frequencies determined. The variation of the values obtained between up- and downsweeps of the magnetic field was used to estimate the uncertainties of the oscillation frequencies and amplitudes. Effective masses were calculated using the LK temperature reduction factor $R_{\rm T}$ described in the main text.  Uncertainties stated in our paper represent the standard error obtained from the fit. Systematic deviations, e.g., arising from the finite field range considered in the analysis, are estimated to be less than $5\%$ of the effective masses inferred from the LK-fits.\\

{\bf Nature and concentration of defects}:
As described in the main text, the magnetic field dependence of the QOs associated with $f_{\alpha}$ and $f_{\beta}$ correspond to Dingle temperatures of $T_{\rm D\alpha, th}\sim1.3\,{\rm K}$ and $T_{\rm D\beta, th}\sim1.2\,{\rm K}$. A related analysis of the anharmonic content of the QOs yield comparable values of $T_{\rm D\alpha, ah}\sim0.9\,{\rm K}$ and $T_{\rm D\beta, ah}\sim0.85\,{\rm K}$. The associated QPL of the order $\tau_{\rm QP}\sim10^{-12}\,{\rm s}$ and Fermi velocity, $v_{\rm F}\sim 3.4*10^5\,{\rm m/s}$, yield mean free paths of the order $l\sim3.4*10^{-7}\,{\rm m}$ (cf. Ref.\,\onlinecite{2022_Guo_NatPhys}). Given that such tiny defect concentrations change the QO spectra significantly, independent microscopic evidence of the nature and concentration of defects is of great interest. 

Unfortunately, there are currently no experimental methods with a suitable sensitivity to provide this information. Adding to the complexity, the QPLs and the Dingle temperatures for different defects may differ by orders of magnitude. Prominent examples reviewed, e.g., in the book of Shoenberg \cite{1984_Shoenberg_Book} and seminal studies of Springford and collaborators \cite{1971_Springford_AdvancesinPhysics, 1977_Paul_JLowTempPhys} include the mechanical state of the sample (strain), the handling of the sample, as well as imperfections such as dislocations and crystalline mosaicity. 

It is, moreover, instructive to comment on the residual resistivity $\rho_0$ as a probe of the purity of bulk materials used widely. First, $\rho_0$ reflects the transport lifetime which is sensitive to large angle scattering. To infer the QPL from the resistivity, hence, details of the scattering processes must be known. Second, different types of defects may be present simultaneously, requiring additional information on their distribution and interplay.  Methods such as remote doping in 2D systems, where the impurities are physically separated from the conduction channel, are not available in bulk materials. In turn, a direct relationship between the residual resistivity and the Dingle temperature does not seem to be possible.

As the Dingle temperature does not distinguish between intra-orbit and inter-orbit contributions, it is essentially not possible to distinguish between different contributions or different types of defects. A possible exception may exist, if the Dingle temperatures of the fundamental orbits are similar. In this case, the effects of intra-orbit scattering cancel in the ratio of the difference frequency with the second harmonic, thus providing a rough estimate of the inter-orbit scattering times a factor reflecting the dimensionality of the system (2D vs 3D, where 3D varies with $\sqrt{B/f}$). 
\\

\textbf{DFT calculations}:
CoSi crystallizes in a simple-cubic lattice (SG\,198) with both atomic species occupying Wyckoff positions $4a$ with $u_\text{Co}=0.143$ and $u_\text{Si}=0.844$ and experimental lattice constant $a=4.444$\,\AA.  The band structure and Fermi surface were calculated with WIEN2k \cite{WIEN2k} and VASP within the generalized gradient approximation by Perdew, Burke and Ernzerhoff \cite{PBE_functional} taking into account SOC. The bands used to generate Fermi surfaces were sampled on a $100\times 100 \times 100$ Monckhorst-Pack grid. Our results are consistent with the reports in the literature \cite{2017_Tang_PhysRevLett, 2018_Pshenay-Severin_JPhysCondensMatter}.
\\\\

\textbf{Predicted QO spectra of $\Gamma$-centered FS sheets}
An apparent question concerns whether the QO frequencies we attribute to the difference and sum of $f_{\alpha}$ and $f_{\beta}$ are, in fact, due other FS sheets. Since CoSi is a semi-metal, only FS sheets at the $\Gamma$ point may be expected apart from those at the R point. 
Unfortunately, the experimental detection of QOs arising from the $\Gamma$-centered FS sheets of CoSi is incomplete. While a frequency smaller than 20\,T has been reported recently that was attributed to a $\Gamma$ pocket\cite{2020_Wang_PhysRevB, Sasmal2022}, the absence of signatures of a larger orbit expected also, may be attributed to its large effective masses in conjunction with the quality of the samples available.

As emphasized in the main text, the observation of the same QO frequencies associated with the R-point in different samples studied by different groups \cite{2022_Huber_PhysRevLett, 2022_Guo_NatPhys, 2019_Rao_Nature, 2022_Guo_NatPhys, 2019_Wu_ChinesePhysLett, 
2019_Xu_PhysRevB, 2020_Wang_PhysRevB, Sasmal2022} establishes the same energy of the electronic bands at the R point. This implies, at high accuracy, a lack of sensitivity of the band structure to defect-induced charge doping. In turn, charge conservation fixes the position of the bands with respect to the Fermi level at the $\Gamma$ point as follows. Shown in Extended Data Fig.\,\ref{fig:EDI4}\,a is the calculated band structure around $\Gamma$. The value of the Fermi energy $E_{\rm F}$ of the band structure as calculated is defined to be zero. To satisfy charge conservation, which requires compensation of the hole pockets at R observed experimentally, the bands at the $\Gamma$ point must be shifted to $+7$\,meV. This adjustment is consistent with the QO oscillation frequency of 20\,T attributed to a $\Gamma$ pocket.\cite{2020_Wang_PhysRevB, Sasmal2022}. It is also consistent with more accurate G$_0$W$_0$-calculations \cite{2018_Pshenay-Severin_JPhysCondensMatter}. 

The FS pockets at the R point observed experimentally in combination with charge conservation fix the Fermi volume at $\Gamma$. Ignoring this strong empirical constraint, charge conservation is strongly violated when shifting the bands by $\pm 20$\,meV with respect to the  point of charge neutrality at $+7$\,meV, as depicted in grey shading. Namely, when shifting the bands in excess of $+20$\,meV the characteristic fourfold degeneracy of all FS sheets at $\Gamma$ shrinks to a point and there are no longer hole-like states. This would imply an electron doping of at least 0.015 electrons per unit cell, which is unrealistically large. Shifting the bands, on the other hand, by the same magnitude of -20\,meV, the resulting position of the Fermi level at -13\-meV would yield an even more dramatic violation of charge neutrality. Namely, for a Fermi value of -1 meV a hole volume of +3\,\% of the BZ is already reached, whereas -13 meV corresponds to a hole volume of +5\,\% and thus an unrealistically large hole-doping of 0.05. 

Marked as colored horizontal lines are the fictitious positions of the Fermi level when shifting the bands such that a QO frequency of $100$\,T frequency may exist at all. 
Two situations may be distinguished. First, when shifting the bands such that $E_{\rm F}=+17.4\,{\rm meV}$, shown in Extended Data Fig.\,\ref{fig:EDI4}\,c, a cross section of $100$\,T exists for field along the $\langle 100 \rangle$ directions. However, the frequency varies strongly as a function of field orientation reaching up to 300\,T. In addition the effective mass varies strongly as a function of field direction.
 Second, when shifting the bands such that $E_{\rm F}=+1.7\,{\rm meV}$, shown in Extended Data Fig.\,\ref{fig:EDI4}\,d, a cross section of $100$\,T exists for field along the $\langle 100 \rangle$ directions that varies strongly as a function of field orientation becoming as low as $\sim 70\,{\rm T}$. Here the effective masses are very large for field along the $\langle 100 \rangle$ directions. Thus, even when strongly violating charge conservation, the FS cross-sections in the vicinity of $\Gamma$ are in stark contrast with experiment.

\textbf{Magnetic breakdown}:
Magnetic breakdown (MB) between SdH orbits around R were evaluated from extremal cross sections of the Fermi surfaces using Chambers' formula for the breakdown field 
\begin{equation}
B_0=\frac{\pi \hbar}{2e} \sqrt{\frac{k_g^3}{a+b}} \mbox{ ,}
\label{eq:chambers}
\end{equation}
where $k_g$ is the gap in $k$-space and $a$ and $b$ are the curvatures of the trajectories at the breakdown junction \cite{1966_Chambers_ProcPhysSoc,1984_Shoenberg_Book}. The MB probability at junction $j$ is then given by $p_j=e^{-\frac{B_0}{B}}$, while the probability for the charge carrier staying on it's trajectory is $q_i=1-p_i$. Only breakdown orbits that are closed after one cycle are considered in the analysis. The junctions are found to be in a regime where breakdown occurs with close to $100$\,\% probability for most of the angular range, consistent with the near-crossings dictated by the quasi-symmetries introduced in Ref.~\cite{2022_Guo_NatPhys}. This results in the simple orbits depicted in grey in Fig.~\ref{fig:2}d--f as discussed below.\\

\textbf{Combined effect of nodal planes and magnetic breakdown}: 

The way in which the remarkably simple spectrum of only two observable main SdH branches arise from the rather intricate FS pockets at the R point has been outlined in Fig.\,1 in the main text and has been clarified in detail in Refs.\,\onlinecite{2022_Huber_PhysRevLett,2022_Guo_NatPhys}. For the sake of a self-contained discussion, we summarize those results here. In particular, we point out that the two approaches in Refs.\,\onlinecite{2022_Huber_PhysRevLett, 2022_Guo_NatPhys, 2022_Wilde_NV} are equivalent as long as the MB gaps are small enough to result in a tunnelling probability close to one.

The FS sheets around the R point arise from four interpenetrating bands as shown in Fig.\,1\,b--f in the main text. In order to understand the extremal orbits one has to consider the band connectivity -- governed by the existence of three mutually perpendicular nodal planes -- on the one hand, and extended regions where the FS sheets are close enough for MB to occur, on the other hand.

In CoSi, all bands are pairwise degenerate on the nodal planes spanning the entire boundary of the simple cubic Brillouin zone. These nodal planes are protected by the joint action of screw rotations and time-reversal symmetry \cite{2022_Huber_PhysRevLett}. The three mutually perpendicular planes intersect at the R point as shown in the upper left corner of Extended Data Fig.\,\ref{fig:EDI3}\,a. The band connectivity due to the nodal planes results in the four FS sheets around R labelled A--D. The FS sheets (C/B) and (A/D) are close to each other in $k$-space in the breakdown regions marked by colored circles (bottom), where almost complete MB occurs. There are no MB junctions between the pairs (A/B) and (C/D), respectively.

The extremal cross sections of these FS sheets are shown in Extended Data Fig.\,\ref{fig:EDI3}\,b for three selected planes. In the (001)-plane, which is a nodal plane, the bands are doubly degenerate. The band structure in this plane thus corresponds exactly to the minimal scenario of a difference frequency of only two almost parallel bands. This results in two SdH orbits (Extended Data Fig.\,\ref{fig:EDI3}\,c) and the occurrence of difference and sum frequencies induced by interband scattering as discussed in the main text. In the (111)-plane and the (110)-plane, four different cross sections exist, with MB junctions marked by colored circles. Because of complete MB, these result in four orbits with pairwise identical cross sections as shown in Extended Data Fig.\,\ref{fig:EDI3}\,c, i.e., for all field directions only two dominant QO frequencies exist.

In Ref.\,\onlinecite{2022_Guo_NatPhys} a slightly different approach was taken. The authors derived an effective perturbation Hamiltonian around the R point, for which the MB regions between FS sheets were degenerate to first order in the new Hamiltonian while small gaps appear only in higher orders of the perturbation. The additional surfaces of degeneracy for this approximate first-order Hamiltonian are shown in green in Extended Data Fig.\,\ref{fig:EDI3}\,d, together with the resulting FS sheets. The corresponding FS sheets are labelled following Ref.\,\onlinecite{2022_Guo_NatPhys} according to the orbital (1/2) and spin (+/-) degree of freedom. These sheets now \emph{intersect} on curves in $k$-space (bottom of Extended Data Fig.\,\ref{fig:EDI3}\,d) that correspond to the breakdown regions in Extended Data Fig.\,\ref{fig:EDI3}\,a. The corresponding extremal cross sections are depicted in Extended Data Fig.\,\ref{fig:EDI3}\,e. Obviously, they result in the same extremal orbits shown in Extended Data Fig.\,\ref{fig:EDI3}\,c without the need to explicitly invoke MB, because the complete MB has already been considered in the construction of the Hamiltonian. The two approaches thus yield consistent results. 

Taken together, the nodal planes and MB regions produce a remarkably simple QO spectrum of only two main frequencies. The difference and sum frequencies then arise due to the nonlinear coupling of orbits on these bands.\\

\textbf{Inconsistencies with text-book mechanisms of combination frequencies}:
Prior to our study four mechanisms have been reported in the literature that may lead to frequencies in quantum oscillatory spectra representing linear combinations of fundamental oscillation frequencies. In CoSi none of these mechanisms allows to account for the properties reported in the manuscript as explained in the following:

\begin{compactitem}
\item[(i)] Magnetic breakdown represents field assisted tunneling of electrons between adjacent orbits in $k$-space. The only case in which MB can lead to frequencies corresponding to the difference of two basis frequencies is a mixed electron-hole orbit~\cite{OBrien2016}. Since the FS pockets around the R point of CoSi are all electron-like \cite{2022_Huber_PhysRevLett,2022_Guo_NatPhys}, MB cannot account for the oscillations at $f_{\beta-\alpha}$ observed experimentally. Furthermore, for a mixed electron-hole orbit one would expect a temperature dependence dictated by the sum of the cyclotron masses of the individual orbits~\cite{2018_vanDelft_PhysRevLett}, also in stark contrast with our experimental observations.

\item[(ii)] Magnetic interactions (MIs) in the context of additional QO frequency contributions refers to the feedback of QOs in the magnetization, i.e., the de Haas-van Alphen effect, on the internal field. MIs may change the QO spectrum and may generate, in principle, a frequency component at $f_{\beta-\alpha}$. A requirement for MIs to become appreciable is a large amplitude $\abs{d\tilde{M}/dB} \geq 1$ \cite{1984_Shoenberg_Book}. We have measured the de Haas-van Alphen oscillations of our CoSi sample, where $\abs{d\tilde{M}/dB}$ is on the order of $10^{-4}$ as shown in Extended Data Fig.\,\ref{fig:EDI6}, i.e., it is orders of magnitude smaller than required for MI. Moreover, oscillation frequencies due to MIs are expected to vanish fast with increasing temperature in stark contrast to our experimental observations. 

\item[(iii)] Quantum oscillations of the chemical potential, also known as chemical potential oscillations (CPOs), arise when the Landau levels are not coupled to a thermodynamic reservoir.  In a multiband metal with several Fermi surface pockets, CPOs may lead to a mixing of different oscillation frequencies~\cite{1997_Alexandrov_PhysicsLettersA}. CPOs are most pronounced for a \mbox{(quasi-)2D} Fermi surface with only a few Landau levels occupied~\cite{1984_Shoenberg_Book}. In turn, CPOs are well-known in dHvA spectra of \mbox{(quasi-)2D} materials. In 3D metals CPOs are suppressed due to the finite background of unquantized electronic states along the field direction and the joint effect of multiple FS sheets, which act as a reservoir. As CoSi hosts multiple 3D Fermi pockets far away from the quantum limit, CPOs will be vanishingly small. In addition, no suppression of the temperature dependence of the oscillations is expected for CPOs in contrast with experiment.

\item[(vi)] Quantum interference oscillations, also known as Stark oscillations (SOs) \cite{1971_Stark_PhysRevLett} may arise from coherent charge carrier pathways between different MB junctions, i.e., they require multiple MB junctions, and MB probabilities $P$ with $P(1-P) \neq 0$. Due to the interference the transmission probability of a quasiparticle between the two points oscillates with the applied magnetic field. In the case of transport properties the allowed oscillation frequencies are given by linear combinations of the frequencies associated with the non-interfering trajectories. They may exhibit a very weak thermal damping~\cite{1983_Kaganov_PhysicsReports}. However, in CoSi for $B \parallel [100]$ no MB junctions exist. For other field directions, where MB junctions are present, the angular evolution of the quantum oscillation amplitudes \cite{2022_Huber_PhysRevLett} and quasi-symmetries \cite{2022_Guo_NatPhys} show that $P \approx 1$ for all field directions and MB junctions. Hence, depending on field orientation the conditions for SOs in CoSi are either not satisfied or SOs are strongly suppressed. 

\end{compactitem}

\subsection{Theoretical framework}

In the following we present details of the theoretical framework underlying the formation of QOs of the QP lifetime. The mechanism we consider is a generalization of so-called magneto-intersubband oscillations in two-dimensional electron gases \cite{Polyanovsky1988, Leadley1992, Coleridge1990, Raikh1994, Goran2009} and quasi-two dimensional metals \cite{Polyanovsky1993,Grigoriev2003, Thomas2008}. A derivation focussing on technical calculations beyond the scope of the work reported here will be presented elsewhere \cite{Leeb2021theory}.

\textbf{Model and conductivity}:
We employ a generic minimal model comprising two rotationally symmetric bands $\lambda = \alpha, \beta$ split in energy. This corresponds to the situation near the R point of CoSi \cite{2022_Huber_PhysRevLett, 2022_Guo_NatPhys}.
In the following section, we set $\hbar = k_B =1$.

In the presence of a magnetic field the electrons are confined to Landau tubes $\epsilon_\lambda = \omega_{c, \lambda} \left(l+\nicefrac{1}{2}\right) + \frac{k_z^2}{2 m_\lambda}-W_\lambda$ where $\omega_{c, \lambda} = \frac{e B}{m_\lambda}$ is the band dependent cyclotron frequency set by the different effective masses $m_\lambda$ and $W_\lambda$ is a relative shift in energy. Note, we use the Landau level dispersion of parabolic bands taking into account the curvature away from the degeneracy point reflecting the experimentally observed band structure. Interestingly, very similar conclusions are obtained for linearly dispersing bands \cite{Leeb2021theory}. Further, we consider randomly distributed short-range impurities by a potential $U(\mathbf{r}) = U_0 \sum_{\mathbf{r}_i} \delta \left(\mathbf{r}-\mathbf{r}_i \right)$ with the generic scattering vertex
\begin{align}
\Lambda = 
\begin{pmatrix}
\Lambda_\alpha & \Lambda_\perp \\
\Lambda_{\perp} & \Lambda_\beta
\end{pmatrix},
\end{align}
where $\Lambda_{\lambda}$ and $\Lambda_\perp$ denote the strength of the intraband and interband scattering, respectively. 

We calculate the conductivity $\hat\sigma_{xx}$ at zero temperature using the Kubo formula \cite{Bastin1971} 
\begin{equation}
\hat\sigma_{xx}(E) = \frac{e^2}{\pi L_x L_y} \mathrm{Tr}_{l,k_x,\lambda,k_z}[v_x \Im G(E) v_x \Im G(E)],
\label{eq:general_Kubo}
\end{equation}
where $G(E)$ is the retarded, impurity averaged Green's function $G_{\lambda,l}(E) = (E-\epsilon_\lambda(l)-\Sigma_\lambda(E))^{-1}$ and $v_x$ is the velocity operator. Note that for short-range impurity scattering, the self-energy $\Sigma_\lambda(E)$ does not depend on the Landau Level index $l$.

Evaluation of all sums results in a generic expression for the conductivity kernel~\cite{Leeb2021theory},
\begin{equation}
\hat \sigma_{xx} = \frac{\sigma_0}{\pi \ell_B} \sum_{\lambda} \frac{\xi_\lambda^\star |\Gamma_\lambda(\xi)|}{1+4\Gamma_\lambda(\xi)^2} \left(\frac{2}{3} \sqrt{2\xi_\lambda^\star} 
+ \sum_{p=1}^\infty \frac{(-1)^p}{\sqrt{p}} \cos\left(2\pi p \xi_\lambda^\star-\frac{\pi}{4}\right) R_\lambda(\xi)^p\right),
\label{eq:3D:cond_res}
\end{equation}
where $\xi_\lambda^\star B$ are the energy dependent frequencies, $\ell_B = \frac{1}{\sqrt{eB}}$ the magnetic length, $\sigma_0 = \nicefrac{2 e^2}{\pi}$ the unit of conductance, and $\Gamma_\lambda = -\Im \nicefrac{\Sigma_\lambda}{\omega_{c, \lambda}}$ the imaginary part of the dimensionless self-energy which can be interpreted as the Landau level broadening (inverse lifetime). In practice, the Landau level broadening is small compared to the cyclotron frequencies and an expansion in $\Gamma_\lambda \propto \nicefrac{1}{\omega_{c\lambda}\tau_\lambda}\ll 1$ recovers the basic formula of the oscillatory component presented in Eq.~\ref{eq:simple_cond}. The associated Dingle factor is given by
\begin{equation}
R_\lambda(\xi) = \exp\left(-2\pi |\Gamma_\lambda(\xi)| \right).
\label{eq:DampinfFac}
\end{equation}

Conventional QO within LK theory are recovered from the above expression when setting $\Im \Sigma_\lambda$ to a constant, the empirical Dingle temperature $\pi T_{D,\lambda}$, such that $R_\lambda(\xi)$ becomes the well known Dingle damping factor~\cite{1984_Shoenberg_Book}
\begin{equation}
R_{D,\lambda} = \exp\left(-2\pi^2 \frac{T_{D,\lambda}}{\omega_{c, \lambda}} \right).
\label{eq:DingleDamping}
\end{equation}
As our main theoretical result, we go beyond this basic approximation and show that the imaginary part of the self-energy, i.e. the band dependent QP lifetime, $\Im \Sigma_\lambda$, exhibits QOs.

\textbf{Calculation of the Self-energy}:
We calculate the self-energy in the self-consistent Born approximation by summing up the diagonal, irreducible Feynman diagrams up to second order
\begin{equation}
\Sigma_\lambda= \begin{tikzpicture}[baseline={(0,-0.07)}]
\draw[dashed] (0,0) -- (0,1);
\fill (0,0) circle (0.07) node[xshift=3pt, anchor=north]{$\Lambda_\lambda$};
\tikzset{shift={(0,1)}}
\tstar{0.03}{0.1}{5}{200}{fill=black}
\tikzset{shift={(1.5,-1)}}
\end{tikzpicture}
+
\begin{tikzpicture}[baseline={(0,-0.07)}]
\draw [-{Stealth[length=3mm, width=2mm]}] (0.75,0) -- (0.87,0) node[anchor=south]{$\lambda$};
\draw (0,0.025) -- (1.5,0.025);
\draw (0,-0.025) -- (1.5,-0.025);
\fill (0,0) circle (0.07) node[xshift=3pt, anchor=north]{$\Lambda_\lambda$};
\fill (1.5,0) circle (0.07)
node[xshift=3pt, anchor=north]{$\Lambda_\lambda$};
\draw[dashed] (0,0) -- +(0.75,1);
\draw[dashed] (1.5,0) -- +(-0.75,1);
\tikzset{shift={(0.75,1)}}
\tstar{0.03}{0.1}{5}{20}{fill=black}
\end{tikzpicture}
+
\begin{tikzpicture}[baseline={(0,-0.07)}]
\draw [-{Stealth[length=3mm, width=2mm]}] (0.75,0) -- (0.87,0) node[anchor=south]{$\bar\lambda$};
\draw (0,0.025) -- (1.5,0.025);
\draw (0,-0.025) -- (1.5,-0.025);
\fill (-0.07,-0.07) rectangle (0.07,0.07)
node[xshift=2pt, yshift=-2pt,anchor=north]{$\Lambda_\perp$};
\fill (1.43,-0.07) rectangle (1.57,0.07)
node[xshift=2pt, yshift=-2pt,anchor=north]{$\Lambda_\perp$};
\draw[dashed] (0,0) -- +(0.75,1);
\draw[dashed] (1.5,0) -- +(-0.75,1);
\tikzset{shift={(0.75,1)}}
\tstar{0.03}{0.1}{5}{20}{fill=black}
\end{tikzpicture}
\end{equation}
where stars denote the impurities and dots and squares denote respectively intraband and interband scattering processes. We note that off-diagonal terms of the self-energy, i.e., processes where the final band index is different from the initial band index, are suppressed by a factor $\nicefrac{T_{D, \lambda}}{W_1-W_2} \ll 1$. The last term in the equation represents the relevant interband scattering process coupling the two bands. One may then derive a self-consistent equation for the imaginary part of the self-energy
\begin{equation}
|\Im \Sigma_\lambda(\xi)| = \pi T_{D,\lambda} \left[1 + \tilde\Lambda_\lambda^2 \sum_{p=1}^\infty \frac{(-1)^p}{\sqrt{2p\xi_\lambda}} \cos\left(2 \pi p \xi^\star_\lambda \right) R_\lambda(\xi)^p  +\tilde\Lambda_{\perp\lambda}^2\sum_{p=1}^\infty \frac{(-1)^p}{\sqrt{2p\xi_{\bar\lambda}}} \cos\left(2 \pi p \xi^\star_{\bar \lambda} \right) R_{\bar\lambda}(\xi)^p\right],
\label{eq:3D:self-energy_selfconsistent1}
\end{equation}
where $\bar \lambda$ represents the other band index with respect to $\lambda$. An analoguous equation exists for the coupled real part of the self-energy~\cite{Leeb2021theory}. Note, as $\Im \Sigma_\lambda \propto \nicefrac{1}{\tau_\lambda}$ we recover Eq.~\eqref{eq:simple_tau} as the lowest order solution of Eq.~\eqref{eq:3D:self-energy_selfconsistent1}.

Crucially, Eq.~\eqref{eq:3D:self-energy_selfconsistent1} reveals that the imaginary part of the self-energy, and hence the QP lifetime, exhibits QOs very similar to the QOs of the conductivity in Eq.~\eqref{eq:3D:cond_res} (the tilde labels renormalized quantities). However, as a truly remarkable facet of Eq.~\eqref{eq:3D:self-energy_selfconsistent1} the self-energy of the electrons of a given band, say $\alpha$, oscillates in addition with the frequency $f_\beta$. These additional oscillations originate from the non-linear interband coupling, and result in the emergence of a sum and difference frequency of the conductivity.

\textbf{QO frequencies and finite temperature behavior}: 
We solve the self-consistent equation for the self-energy Eq.~\eqref{eq:3D:self-energy_selfconsistent1} by iterative insertion and expansion in orders of the Dingle factor $R_{D,\lambda}$, keeping terms up to the second harmonics. Substituting the result into Eq.~\eqref{eq:3D:cond_res}, we obtain the conductivity kernel $\hat \sigma_{xx}$. The resulting sum consists of summands $\propto \cos \left(2 \pi \frac{f(\xi)}{B} \right)$ where $f = f_\alpha$, $f_\beta$, $2 f_\alpha$, $2 f_\beta$, $f_\alpha+f_\beta$, $f_\beta-f_\alpha$ and more frequencies beyond the second order. Physically, the oscillations with $f_\alpha+f_\beta$ and $f_\beta-f_\alpha$ reflect the interference of the oscillations of the lifetime with the intrinsic oscillations of the conductivity. 

We determine the conductivity $\sigma_{xx}$ at finite temperatures from the conductivity kernel $\hat \sigma_{xx}$ by a convolution with the derivative of the Fermi distribution function\cite{Leeb2021theory}. The final result reads
\begin{align}
\frac{\sigma_{xx}}{\sigma_0} = &\sum_\lambda A_{\lambda} \cos \left(2 \pi\frac{f_\lambda}{B}-\frac{\pi}{4}\right)  R_{D,\lambda} R_T\left(m_\lambda\right) 
+ \sum_\lambda A_{2\lambda} \cos \left(4 \pi\frac{f_\lambda}{B}-\frac{\pi}{4}\right)  R_{D,\lambda}^2 R_T\left(2 m_\lambda \right)
\nonumber \\
&+ A_{\alpha+\beta} \cos \left(2 \pi\frac{f_\alpha+f_\beta}{B}-\frac{\pi}{2}\right)  R_{D,\alpha} R_{D,\beta} R_T\left(m_\alpha + m_\beta\right)
\nonumber \\
&+ A_{\beta-\alpha} \cos \left(2 \pi\frac{f_\beta-f_\alpha}{B}\right)  R_{D,\alpha} R_{D, \beta} R_T\left(m_\beta-m_\alpha\right),
\label{app:3D:cond_final}
\end{align}
where the amplitudes $A_i$ only influence the QOs weakly \cite{Leeb2021theory}. Importantly, we find that $A_{\alpha \pm \beta}/A_{\lambda} \approx \nicefrac{\Lambda_\perp}{\Lambda_\lambda}\left(\nicefrac{1}{\sqrt{f_\alpha}}+\nicefrac{1}{\sqrt{f_\beta}}\right)$, leading to the conclusion that even if the amplitude of the difference frequency is small interband coupling can be strong, see Fig.~\ref{fig:EDI5}. Note that the difference and sum of the frequencies are damped with the sum of the Dingle temperatures of the fundamental oscillation frequencies. In contrast, the decay in temperature, set by $R_T$, is governed by the difference or the sum of the effective masses of the fundamental frequencies. Furthermore, the sum and the difference frequency inherit the phases of the fundamental frequencies, i.e., the phases of the combination frequencies are $\varphi_{\beta\pm\alpha} = \varphi_\beta \pm \varphi_\alpha$, which provides another experimental test permitting their identification.

\textbf{Interband coupling from magnetic impurity scattering:}
The interband impurity scattering contribution is key for observing difference frequency QOs. In the following, we evaluate how the underlying spin-1 multifold fermion excitation may be responsible for the strong interband impurity scattering channels. CoSi belongs to SG 198 for which excitations around R can be described by a generic low-energy $\mathbf{k}\cdot\mathbf{p}$-theory \cite{Manes2012,Bradlyn2016,2018_Pshenay-Severin_JPhysCondensMatter}. The band structure features a spin-1 multifold fermion which forms out of two Weyl nodes due to spin-orbit coupling. The $\mathbf{k}\cdot\mathbf{p}$-Hamiltonian ($\mathbf{k}$ is the momentum with respect to the R point) reads
\begin{equation}
H_R = \nu \mathbbm{1} \otimes \mathbf{k} \cdot \boldsymbol{\sigma} + \frac{\Delta}{4}\left(\sigma_x \otimes \sigma_x + \sigma_y \otimes \sigma_y + \sigma_z \otimes \sigma_z\right),
\label{eq:H_manes}
\end{equation}
where $\sigma_i$ are the Pauli matrices and for CoSi $\nu \approx -\unit[1.3]{eV}$ and $\Delta \approx \unit[31]{meV}$, is known to be a good effective description along the surface of the Brillouin zone \cite{2018_Pshenay-Severin_JPhysCondensMatter}. Eq.~\eqref{eq:H_manes} is denoted in the orbital basis where the first index is spin $s$ and the second is pseudospin $\tau$ related to the sublattice structure. We denote the bands $E_{\alpha_{\v{k}}} = \nu |\v{k}| + \frac{\Delta}{4}$, $E_{\beta_{\v{k}}} = -\frac{\Delta}{4} + \sqrt{\left(\frac{\Delta}{2}\right)^2+\nu^2 \v{k}^2}$ of the low-energy Hamiltonian by $\ket{\lambda}$ where $\ket{\alpha_{\v{k}}}$ and $\ket{\beta_{\v{k}}}$ are the two relevant bands for the observed QOs, as they set the frequencies (see Extended Data Fig.\,\ref{fig:EDI5}). Importantly, the electron spin in the band $\alpha$ ($\beta$) is polarized antiparallel (parallel) to its momentum, consistent with CoSi \cite{2022_Guo_NatPhys}.

We consider simple potential impurities which are distributed uniformly over the system. The simplest and most common form of impurity potentials is diagonal in the orbital basis  $\Lambda = \sum_{s, \tau} \Lambda_{s,\tau} \ket{s,\tau} \bra{s,\tau}$ but spin and pseudospin dependent. By using the analytic expressions for the band eigenstates, we can then show that the interband coupling 
$\Lambda_\perp = \bra{\alpha_{\v{k}}}\Lambda \ket{\beta_{\v{k}'}} \propto \sum_\tau \left(\Lambda_{\downarrow, \tau}-\Lambda_{\uparrow, \tau}\right)$ is significant if the scattering breaks time reversal symmetry (TRS).

In the case of CoSi, it has been known for a long time that impurities lead to local moment formation on the Co atoms, e.g. for Fe impurities in the crystal \cite{Yasuoka1974,Wernick1972,Wertheim1966,Kawarazaki1972}. These TRS breaking impurities then directly imply different signs for $\Lambda_{\uparrow,\tau}$ and $\Lambda_{\downarrow,\tau}$, hence, they already lead at the basic level of potential impurity scattering in the orbital basis to a finite interband coupling.

Note that the fact that the amplitudes of the difference and sum frequency are much smaller than the amplitudes of the ordinary second harmonics does not imply that $\Lambda_\perp \ll \Lambda_i$. QOs of the lifetime come with a different prefactor leading to a smaller contribution to the total spectrum in 3D. In fact the measured QO spectrum in CoSi suggests that interband and intraband scattering are roughly equally strong, increasing the importance of the interband scattering mechanism.

Taken together, we argue that the universal band structure related to the multifold spin-1 fermions in CoSi together with magnetic impurities may give rise to an effective interband coupling that accounts for the observation of QOs of the QP lifetime.


\newpage

\newpage
\section*{Acknowledgements}

We wish to thank A. Schnyder for discussions. J.K. acknowledges helpful discussions with N.R. Cooper. V.L. acknowledges support from the Studienstiftung des deutschen Volkes. N.H. and V.L. acknowledge support from the TUM Graduate School. M.A.W. and C.P. acknowledge support through DFG TRR80 (project-id 107745057) and  DFG GACR Projekt WI 3320/3-1, C.P. acknowledges support through DFG SPP 2137 (Skyrmionics) under grant number PF393/19 (project-id 403191981), ERC Advanced Grant No. 788031 (ExQuiSid), and Germany's excellence strategy under EXC-2111 390814868. J. K. acknowledges support from the Imperial-TUM flagship partnership. This research is part of the Munich Quantum Valley, which is supported by the Bavarian state government with funds from the Hightech Agenda Bayern Plus. 


\section*{Author Contributions}
M.A.W. and C.P. conceived and started this study.
M.A.W., C.P., and J.K. proposed the interpretation.
G.B. and A.B. prepared and characterized the samples. 
N.H. conducted the measurements and analyzed the data. 
M.A.W. and V.L. performed band structure calculations. 
M.A.W. and N.H. connected the experimental data with the calculated band structure.  
V.L. and J.K. developed the theoretical analysis. 
C.P., M.A.W., and J.K. wrote the manuscript with contributions from N.H. and V.L..
All authors discussed the data and commented on the manuscript.





\section*{Data availability}
Data reported in this paper are available at:
https://doi.org/10.5281/zenodo.7957067


\section*{Competing interests}
The authors declare no competing interests.


\section*{Correspondence and requests for materials}
Correspondence and requests for materials should be addressed to M.A.W, C.P. or J.K.


\newpage

\section*{Extended Data}

\setcounter{figure}{0}
\captionsetup[figure]{labelfont={bf},name={Extended Data Fig.},labelsep=space}

\setcounter{table}{0}
\captionsetup[table]{labelfont={bf},name={Extended Data Table},labelsep=space}


\clearpage \thispagestyle{empty}

\begin{figure*}[h]
	\centerline{\includegraphics[width=0.8\textwidth,clip=]{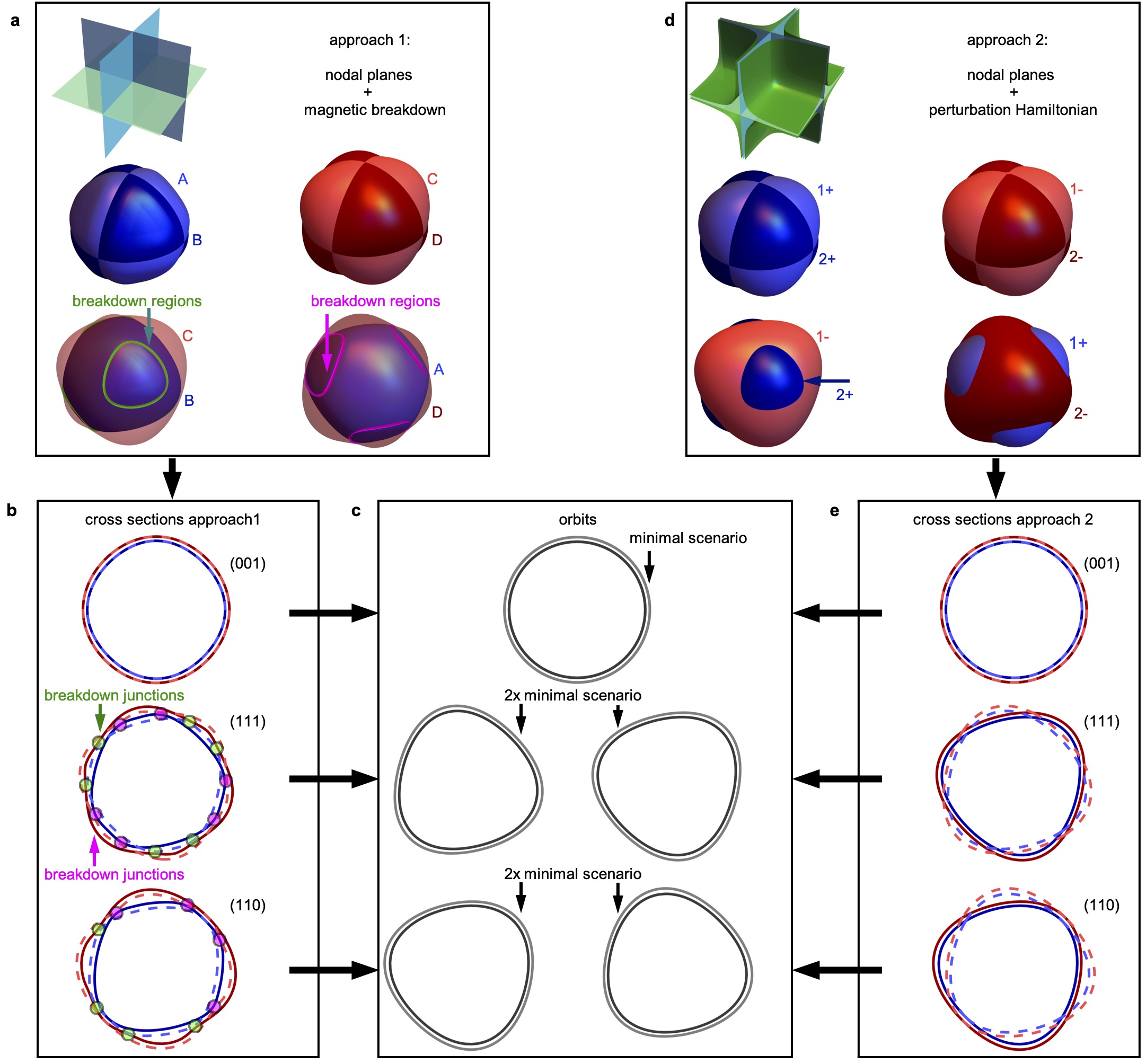}}
\linespread{1.0}\selectfont{}
\caption{\raggedright
{\bf $|$ Equivalence of predicted quantum oscillation orbits when invoking magnetic
breakdown (Ref.\,\onlinecite{2022_Huber_PhysRevLett}) or hidden quasi-symmetries (Ref.\,\onlinecite{2022_Guo_NatPhys}).} Panels are organized in the spirit of a flow-chart.
{\bf a}, Approach 1 comprises nodal planes and magnetic breakdown (MB) \cite{2022_Huber_PhysRevLett}. Three mutually perpendicular nodal planes at the R point as depicted at the top of the panel enforce pairwise band degeneracies of the FS sheets labelled A, B and C, D. Numerically calculated MB with probabilities close to 1 occur between sheet pairs (C,B) and (A,D) along the lines as marked.
{\bf b}, Extremal cross section for three selected planes. In the (001)-plane, the cross sections are doubly degenerate. In the (111)-plane and the (110)-plane, four different cross sections exist, with MB junctions denoted by colored circles. The cross-sectional areas are always pairwise identical.
{\bf c}, Effective extremal orbits corresponding to panels b and e. For all field directions two dominant QO frequencies exist.
{\bf d}, Approach 2 comprises nodal planes and hidden quasi-symmetries \cite{2022_Guo_NatPhys}. Three mutually perpendicular nodal planes at the R point are enclosed by regimes of hidden quasi-symmetry up to the enclosing surfaces (green) as depicted at the top of the panel. Within the green regimes near degeneracies are approximated by exact degeneracies in first-order perturbation theory. The FS sheets of the perturbation Hamiltonian are labelled as orbital $(1 / 2)$ and spin $( +/ -)$ degrees of freedom, where sheet pairs $(1- / 2+ )$ and $(1+/ 2-)$ intersect at the surfaces of the quasi-symmetry.
{\bf e}, Extremal cross sections for the same three selected planes shown in b. In the (001)-plane the cross-sections are doubly degenerate identical to b. In the (111)-plane and the (110)-plane, the cross sections correspond to the extremal orbits as depicted in b with breakdown gaps that approach zero. Taken together approaches 1 and 2 yield the same orbits as depicted in c and thus the same two QO frequencies.
}
\label{fig:EDI3}
\end{figure*}

\clearpage \thispagestyle{empty}

\begin{figure*}[h]
	\centerline{\includegraphics[width=1.0\textwidth,clip=]{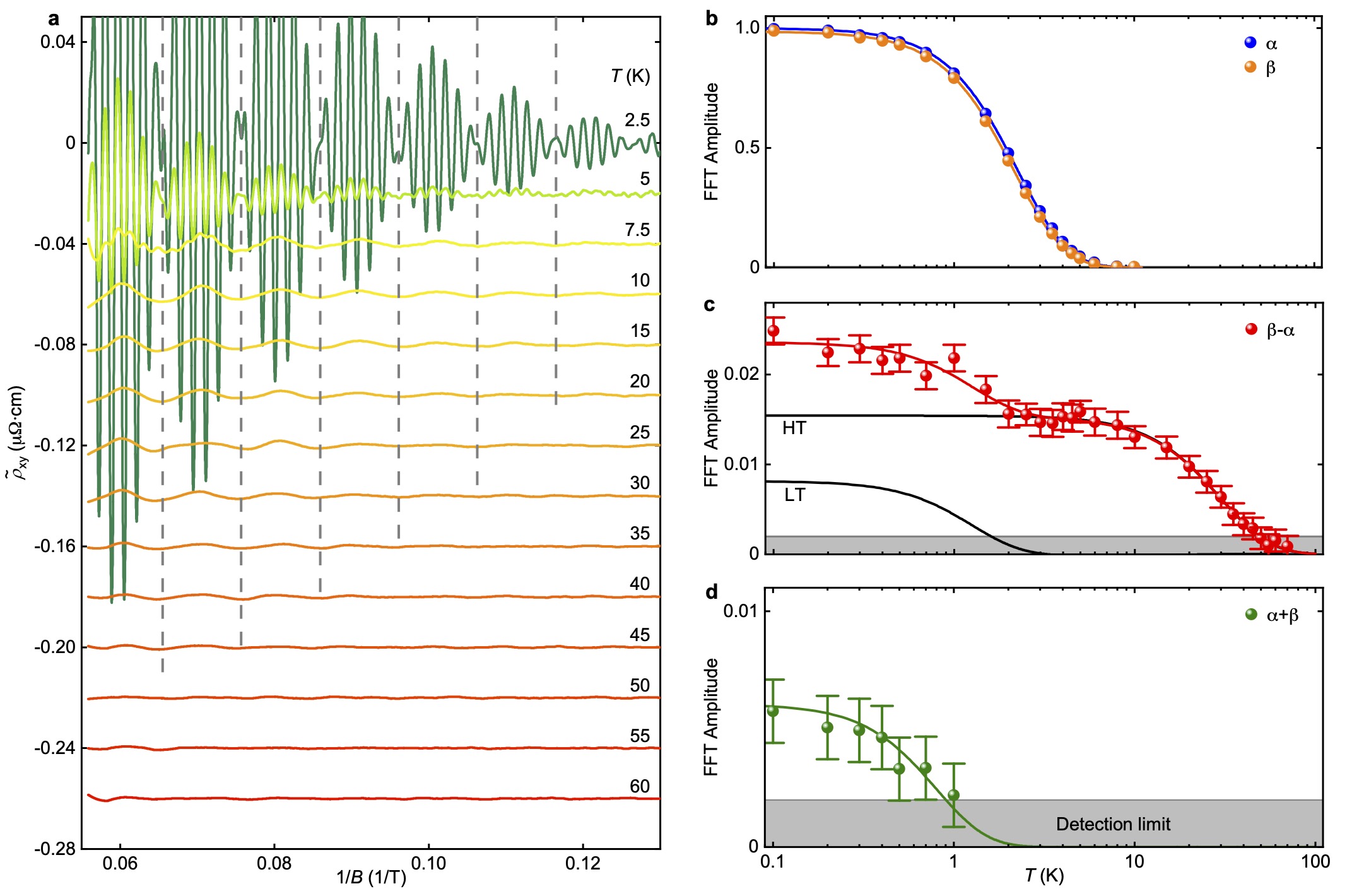}}
\linespread{1.0}\selectfont{}
\caption{\raggedright
{\bf $|$ Temperature dependence of Shubnikov-de Haas oscillations in the Hall resistivity of CoSi.} 
Oscillation amplitudes shown in panels b-d were analyzed in the magnetic field range between 9 and 18\,T with $B$ applied along the [001] direction and normalized to the amplitude of $f_{\alpha}$ at $T=20$\,mK. 
{\bf a}, Oscillatory component of \ryx{} as a function of inverse magnetic field at different temperatures. Curves are shifted by a constant.
{\bf b},~Oscillation amplitudes of the $\alpha$ and $\beta$ frequencies. Lines represent a fit with the temperature reduction factor $R_{\rm T}$ in the Lifshitz--Kosevich formalism. Effective masses inferred from the fit are $m_{\alpha}^{*}=(0.92 \pm 0.01)\,m_\mathrm{e}$ and $m_{\beta}^{*}=(0.95 \pm 0.01)\,m_\mathrm{e}$. Error bars are smaller than the data points. {\bf c},~Oscillation amplitudes of the difference frequency  $f_{\beta-\alpha}$. The red line represents a two-component fit of the temperature reduction factor yielding effective masses of $(0.07 \pm 0.01)\,m_\mathrm{e}$ and $(1.6 \pm 0.3)\,m_\mathrm{e}$, corresponding to the difference and the sum of the individual masses within the error bars and consistent with the quasiparticle lifetime oscillations reported in this work. The black lines represent the individual components of the fit, where HT and LT denote the high and low temperature behavior, respectively.
{\bf d},~Amplitude of the sum frequency $f_{\alpha+\beta}$. The line represents a LK fit yielding a cyclotron mass of $(2.6 \pm 0.4)\,m_\mathrm{e}$. 
}
\label{fig:EDI1}
\end{figure*}

\clearpage \thispagestyle{empty}

\begin{figure*}[h]
	\centerline{\includegraphics[width=0.9\textwidth,clip=]{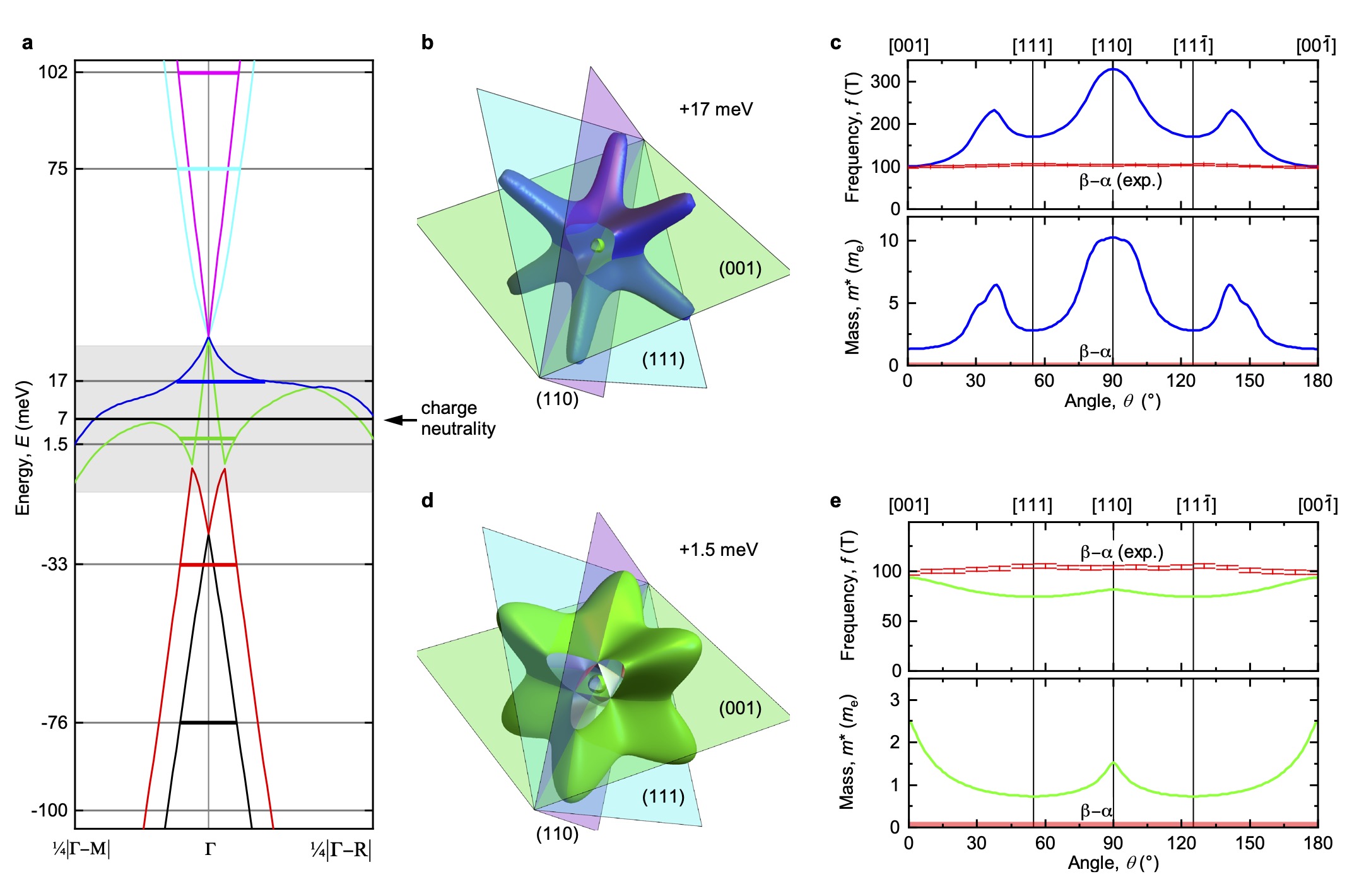}}
\linespread{1.0}\selectfont{}
\caption{\raggedright
{\bf $|$ 
Calculated FS orbits near the $\Gamma$ point of CoSi under various conditions.}
The observation of essentially identical values of $f_{\alpha}$ and $f_{\beta}$ in different samples studied by different groups \cite{2019_Wu_ChinesePhysLett, 2019_Xu_PhysRevB, 2020_Wang_PhysRevB, 2022_Huber_PhysRevLett, 2022_Guo_NatPhys, Sasmal2022} establishes the same energy of the electronic bands at the R point and thus a lack of sensitivity of the band structure to defect-induced charge doping. In turn, charge conservation enforces a position of the bands with respect to the Fermi level at the $\Gamma$ point within one meV. No QO frequencies $\sim 100\,{\rm T}$ are expected under these conditions. 
{\bf a}, Calculated band structure in the vicinity of the $\Gamma$ point of CoSi. For the FS pockets at the R point observed experimentally, a value of $E_{\rm F}=(+7\,\pm1){\rm meV}$ at the $\Gamma$ point, marked by the black horizontal line, corresponds to charge neutrality. 
The grey shading depicts a range of $\pm20\,{\rm meV}$ beyond which hole carriers vanish altogether and charge conservation would be violated dramatically (see methods for details). Colored horizontal lines at $E_{\rm F}=+17\,{\rm meV}$ and $E_{\rm F}=+1.5\,{\rm meV}$ denote locations, where FS cross sections with QO frequencies of $\sim100\,{\rm T}$ are expected, however, for $B \parallel \langle 001 \rangle$ only. 
{\bf b}, FS sheets centered at $\Gamma$ for $E_{\rm F}=+17\,{\rm meV}$, where the sheets depicted in blue and green shading correspond to the blue and green bands shown in panel a. 
{\bf c}, Upper panel: Calculated angular dispersion $f(\theta)$ of the blue band for $E_{\rm F}=+17\,{\rm meV}$, as compared to the experimental data (red symbols) shown in Fig.\,\ref{fig:3}i. Lower panel: Calculated dispersion of the cyclotron mass (blue) as compared to the experimental value (red), where strong disagreement would be expected. QO with a frequency $\sim100\,{\rm T}$ are expected for $B \parallel \langle 001 \rangle$ only, where the predicted mass differs strongly from experiment.
{\bf d}, Green FS sheet centered at $\Gamma$ for $E_{\rm F}=+1.5\,{\rm meV}$. 
{\bf e}, Upper panel: Calculated angular dispersion $f(\theta)$ of the green band for $E_{\rm F}=+1.5\,{\rm meV}$, as compared to the experimental data (red symbols) shown in Fig.\,\ref{fig:3}i. Lower panel: Calculated dispersion of the cyclotron mass (green) as compared to the experimental value (red), where strong disagreement would be expected. QO with a frequency $\sim100\,{\rm T}$ are expected for $B \parallel \langle 001 \rangle$ only, where the predicted mass differs strongly from experiment.
}
\label{fig:EDI4}
\end{figure*}

\clearpage \thispagestyle{empty}

\begin{figure*}[h]
	\centerline{\includegraphics[width=1.0\textwidth,clip=]{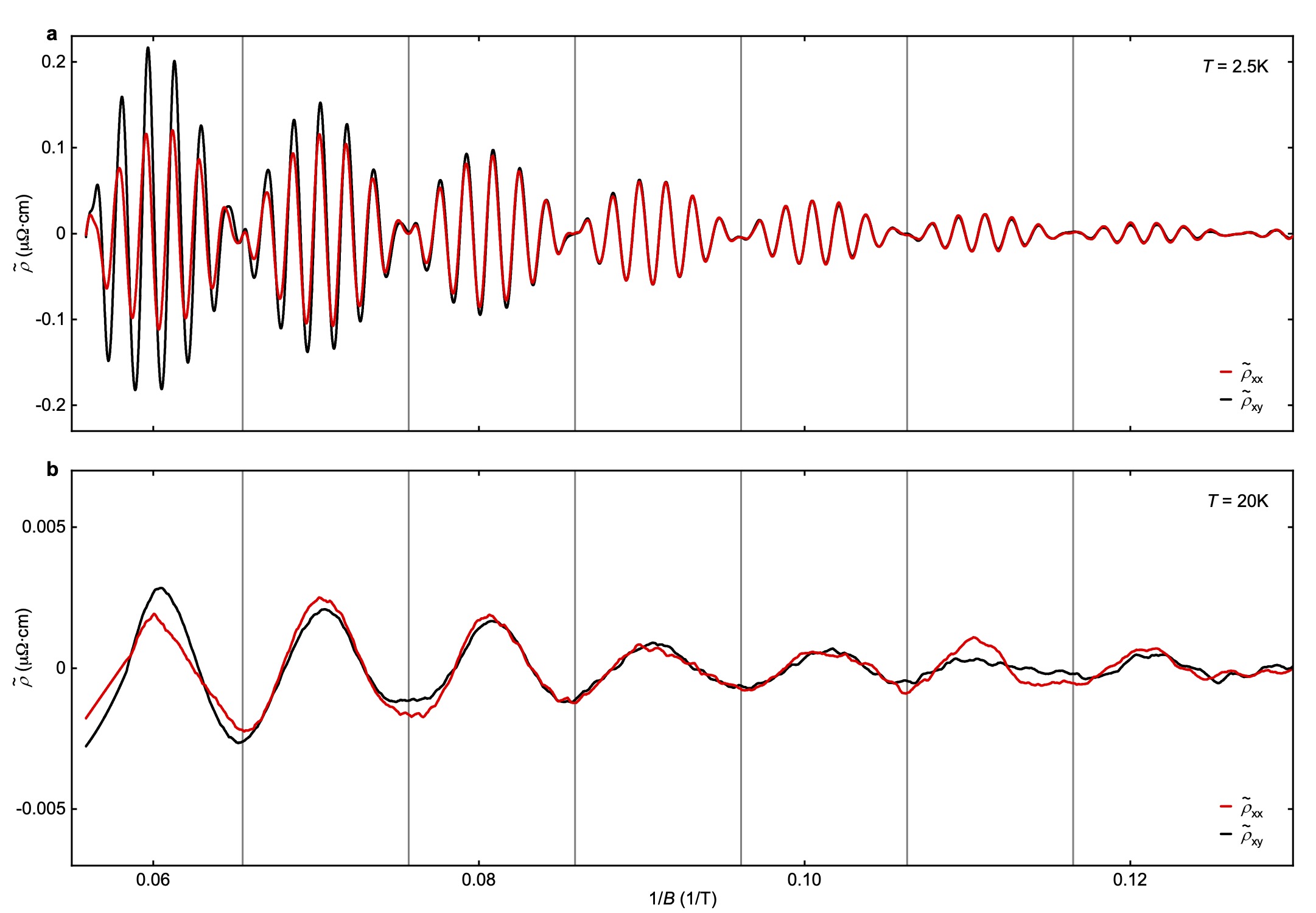}}
\linespread{1.0}\selectfont{}
\caption{\raggedright
{\bf $|$ Analysis of the phase relation between the SdH oscillations at different temperatures.}
{\bf a}, Oscillatory component of the transverse magnetoresistivity, \tilrxx{}, and Hall resistivity, \tilryx{}, as a function of inverse magnetic field at $T=2.5$\,K. A pronounced beating pattern originates from oscillations at $f_\alpha$ and $f_\beta$. Nodes of the beating pattern are indicated by vertical lines. The envelope of the beating pattern may be expressed as 
$ \cos\left(2\pi (f_{\beta}-f_{\alpha})/(2 B) + (\varphi_{\beta}-\varphi_{\alpha})/2\right) $ 
where $\varphi_{\alpha}$ and $\varphi_{\beta}$ are the phases of $f_\alpha$ and $f_\beta$, respectively. An analysis using the frequencies determined from the FFT peaks, notably $f_{\alpha}=565\,T$ and $f_{\beta}=663\,T$, yields a phase difference of $\varphi_{\beta}-\varphi_{\alpha} = 0.16\pi$. We note that this value sensitively depends on the precise value of the frequency difference $f_{\beta}-f_{\alpha}$.
{\bf b}, Oscillatory component of the resistivities as a function of inverse applied magnetic field at $T=20$\,K. The oscillations at $f_{\alpha}$ and $f_{\beta}$ are strongly suppressed at this temperature. Only the slow oscillations at $f_{\beta-\alpha}$ are visible. Here, minima of the oscillations coincide with the nodes of the beating pattern shown in panel {\bf a}. Neglecting the amplitude, the oscillation at $f_{\beta-\alpha}$ may be described by a term reading
$ \cos\left(2\pi f_{\beta-\alpha}/B + \varphi_{\beta-\alpha}\right)$, which oscillates with the same frequency as the nodes in the beating pattern. The phase $\varphi_{\beta-\alpha}$ matches the phase difference $\varphi_{\beta}-\phi_{\alpha}$.
}
\label{fig:EDI2}
\end{figure*}

\clearpage \thispagestyle{empty}

\begin{figure*}[h]
	\centerline{\includegraphics[width=0.8\textwidth,clip=]{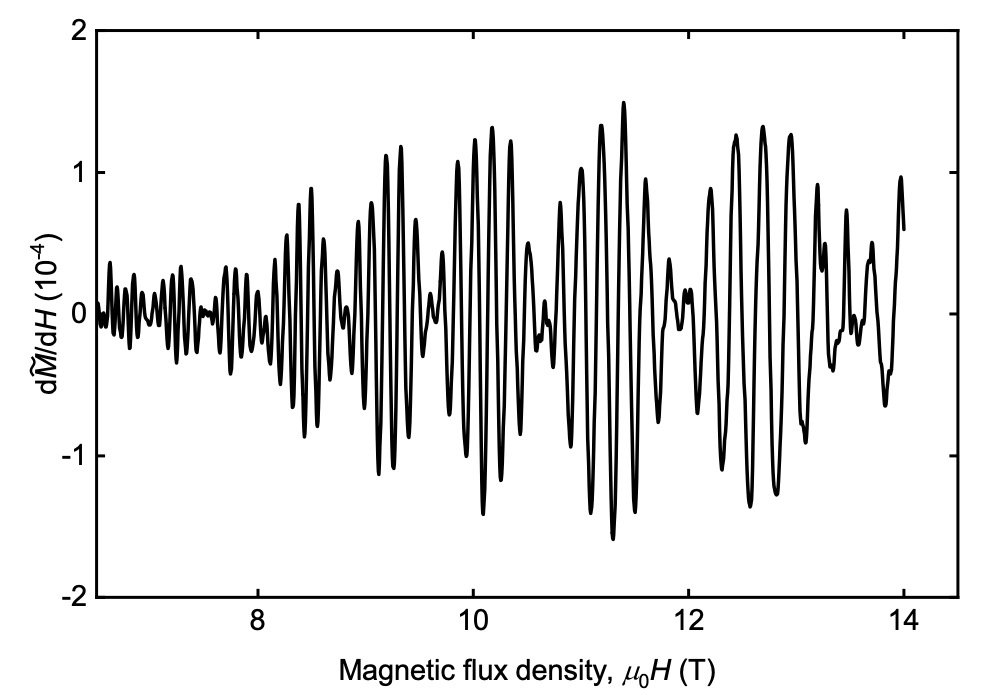}}
\linespread{1.0}\selectfont{}
\caption{\raggedright
{\bf $|$ Quantitative assessment of the effects of magnetic interactions in CoSi.} Changes of the magnetization due to de Haas-van Alphen oscillations may generate variations of the internal field that result in oscillatory signal components observed in the physical properties. This feedback is known as magnetic interaction. 
Shown is the derivative of the oscillatory part of the measured magnetization in SI units as a function of applied magnetic field. For contributions of the order $d\tilde{M}/dH\sim 1$, oscillatory signal contributions due to magnetic interactions are commonly expected \cite{1984_Shoenberg_Book}. The de Haas-van Alphen contribution observed experimentally is well below this limit. Numerical simulations of the effect of magnetic interactions expected in our study establish that the amplitudes of the de Haas-van Alphen oscillations at $f_{\alpha}$ and $f_{\beta}$ are several orders of magnitude below the value required to account for the oscillatory signal components at the difference of the frequencies we observed experimentally.
}
\label{fig:EDI6}
\end{figure*}

\clearpage \thispagestyle{empty}

\begin{figure*}[h]
	\centerline{\includegraphics[scale=0.8,clip=]{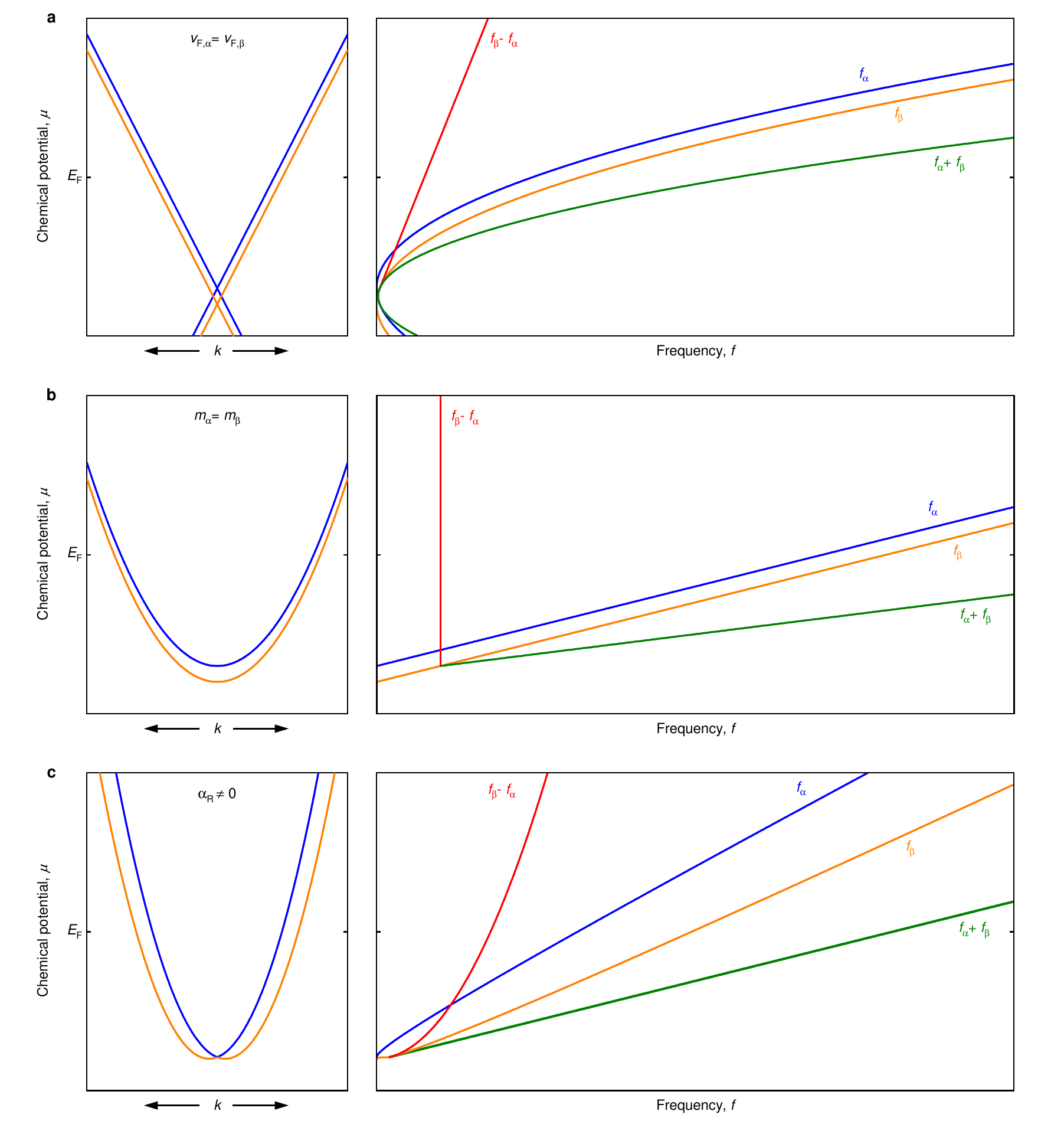}}
\linespread{1.0}\selectfont{}
\caption{\raggedright
{\bf $|$ Characteristic frequencies of quantum oscillations of the quasiparticle lifetime for generic types of band pairs.}
Precondition for QO of the QPL is a general form of interaction such as scattering from defects or collective excitations that generate a non-linear intra-orbit or inter-orbit coupling.
{\bf a}, Left: Two linear bands with equal Fermi velocities, $v_{\rm F,\alpha}=v_{\rm F,\beta}$, and a small offset. This configuration may be expected near multi-fold band crossings as in CoSi. The crossings need not be topological. Right: The small variation of $f_\beta-f_\alpha$ due to the band offset leads to a strongly reduced suppression of the dephasing of the oscillation amplitude as a function of temperature. 
{\bf b}, Left: Parabolic bands with equal masses, $m_\alpha=m_\beta$, and a small offset. Right: The difference frequency $f_\beta-f_\alpha$ is independent of the Fermi level as the cyclotron masses are equal. The dephasing as a function of temperature and hence the decay of the oscillation amplitude with increasing temperature is suppressed completely.
{\bf c}, Left: Rashba-type spin-orbit splitting, where the Rashba parameter $\alpha_{\mathrm{R}} \neq 0$. Right: The temperature dephasing depends on $E_{\rm F}$, since the slope of $f_\beta-f_\alpha$ is not constant.
}
\label{fig:EDI7}
\end{figure*}

\clearpage \thispagestyle{empty}

\begin{figure*}[h]
	\centerline{\includegraphics[scale=1,clip=]{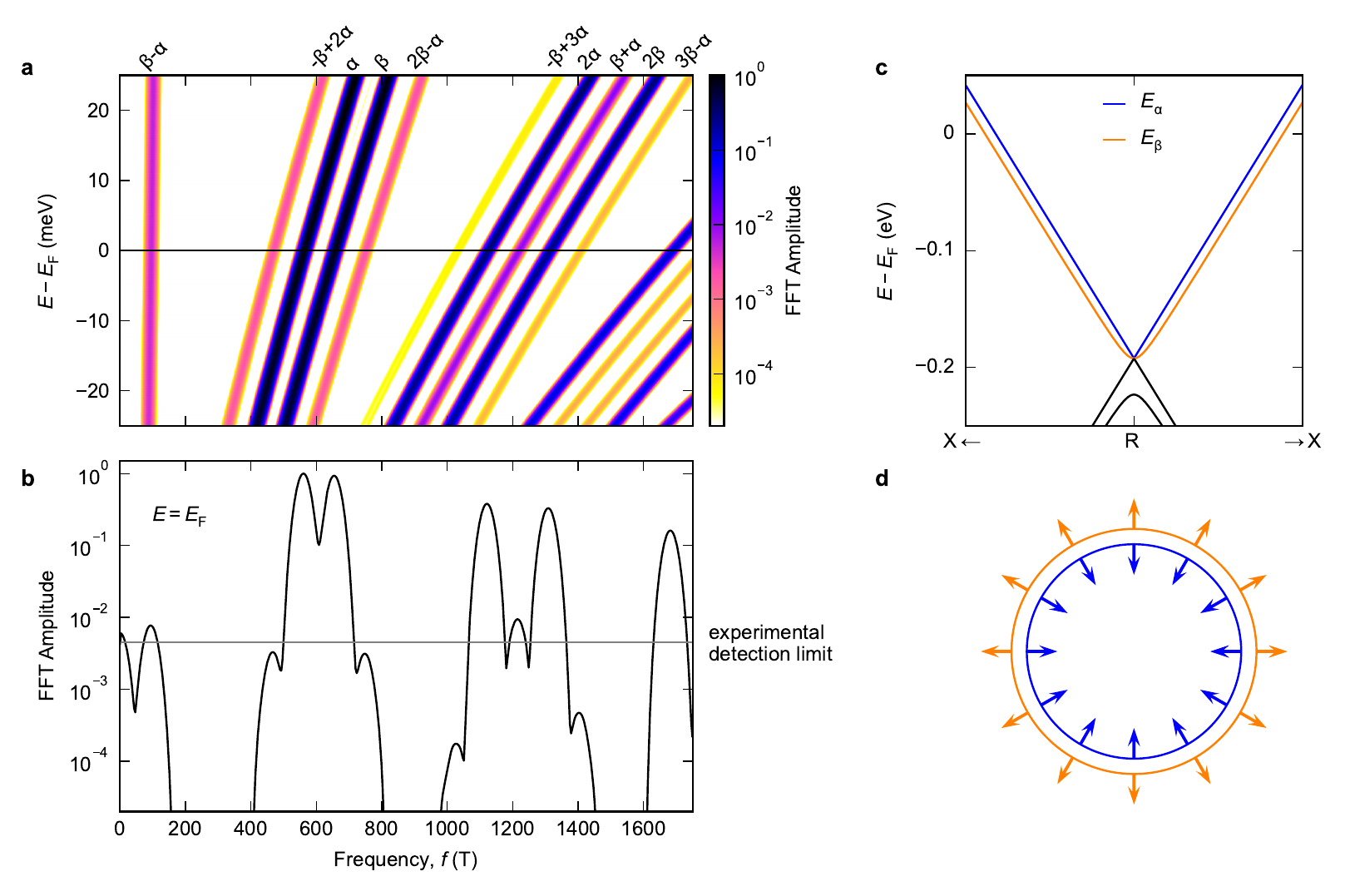}}
\linespread{1.0}\selectfont{}
\caption{\raggedright
{\bf $|$ Predicted QO spectra and possible mechanism of interband scattering.} 
The dependence of the QO frequencies on energy is key to understanding their temperature dependence. {\bf a}, FFT amplitude of Eq.~\ref{app:3D:cond_final} over an extended $B$-field range at $T=0$. In addition to Eq.~\ref{app:3D:cond_final}, leading contributions up to the fourth order in the Dingle factor are included which can be evaluated systematically. Experimental values for $m_\lambda^*$ and $T_{D, \lambda} \approx \unit[0.5]{K}$ were used consistent with CoSi, as well as $f_\alpha(E),f_\beta(E)$ from the DFT calculations. {\bf b}, Cut along $E=E_F$ for CoSi. This may be compared to the spectrum recorded experimentally. Peaks shown in panels {\bf a} and {\bf b} appear broad due to the $\log$-scale. {\bf c}, Band structure of the effective $\mathbf{k} \cdot \mathbf{p}$-Hamiltonian around the R point Eq.~\ref{eq:H_manes} used to model a possible mechanism for non-linear interband scattering. 
{\bf d}, Schematic Fermi surfaces in the effective $\v{k} \cdot \v{p}$-Hamiltonian (projection on the $k_z=0$ plane with respect to R) in which the spin is polarized perpendicular to the Fermi surface and orientated in opposing directions on the two bands. This is consistent with results obtained in DFT calculations \cite{2022_Guo_NatPhys}.
}
\label{fig:EDI5}
\end{figure*}

\end{document}